\documentclass[longauth]{aa}

\usepackage{graphicx,latexsym,amssymb,times}
\usepackage{lscape}
\usepackage{multirow}
\usepackage{natbib}
\usepackage{color}
\bibpunct{(}{)}{;}{a}{}{,} 

\begin{document}
\newcommand {\sax} {{\it Beppo}SAX }
\newcommand {\asca} {{\it ASCA} }
\newcommand {\xmm } {{\it XMM-Newton}}
\newcommand {\rosat} {{\it ROSAT }}
\newcommand {\rxte} {{\it RXTE} }
\newcommand {\swift} {{\it Swift}}
\newcommand {{\chandra}}{{\it Chandra}}
\newcommand {\hess} {{HESS} }
\newcommand {\rchisq} {$\chi_{\nu} ^{2}$} 
\newcommand {\chisq} {$\chi^{2}$}
\newcommand {\ergs}[1]{$\times10^{#1}$ erg cm$^{-2}$ s$^{-1}$}
\newcommand {\e}[1]{$\;\times10^{#1}$}
\newcommand {\nufnu}{$\nu F_{\nu}$ }
\newcommand {\nw}{nW m$^{-2}$sr$^{-1}$}
\newcommand {\micron}{$\mu$m }
\newcommand {\cms}{cm$^{-2}$s$^{-1}$ }
\newcommand {\mjd}{MJD$^\star$}
\newcommand {\glast} {{\it Fermi}-LAT}

\title{Simultaneous multiwavelength observations of the second exceptional
       $\gamma$-ray flare of PKS\,2155--304 in July 2006}


\author{F. Aharonian\inst{1,13}
 \and A.G.~Akhperjanian \inst{2}
 \and G.~Anton \inst{16}
 \and U.~Barres de Almeida \inst{8} \thanks{supported by CAPES Foundation, Ministry of Education of Brazil}
 \and A.R.~Bazer-Bachi \inst{3}
 \and Y.~Becherini \inst{12}
 \and B.~Behera \inst{14}
 \and W.~Benbow \inst{1}
 \and K.~Bernl\"ohr \inst{1,5}
 \and C.~Boisson \inst{6}
% ...
}

\offprints{luigi.costamante@stanford.edu or Rolf.Buehler@mpi-hd.mpg.de}
\institute{
Max-Planck-Institut f\"ur Kernphysik, P.O. Box 103980, D 69029
Heidelberg, Germany
\and
 Yerevan Physics Institute, 2 Alikhanian Brothers St., 375036 Yerevan,
Armenia
\and
Centre d'Etude Spatiale des Rayonnements, CNRS/UPS, 9 av. du Colonel Roche, BP
4346, F-31029 Toulouse Cedex 4, France
\and
Universit\"at Hamburg, Institut f\"ur Experimentalphysik, Luruper Chaussee
149, D 22761 Hamburg, Germany
\and
Institut f\"ur Physik, Humboldt-Universit\"at zu Berlin, Newtonstr. 15,
D 12489 Berlin, Germany
\and
LUTH, Observatoire de Paris, CNRS, Universit\'e Paris Diderot, 5 Place Jules Janssen, 92190 Meudon, 
France
\and
IRFU/DSM/CEA, CE Saclay, F-91191
Gif-sur-Yvette, Cedex, France
\and
University of Durham, Department of Physics, South Road, Durham DH1 3LE,
U.K.
\and
Unit for Space Physics, North-West University, Potchefstroom 2520,
    South Africa
\and
Laboratoire Leprince-Ringuet, Ecole Polytechnique, CNRS/IN2P3,
 F-91128 Palaiseau, France
\and 
Laboratoire d'Annecy-le-Vieux de Physique des Particules, CNRS/IN2P3,
9 Chemin de Bellevue - BP 110 F-74941 Annecy-le-Vieux Cedex, France
\and
Astroparticule et Cosmologie (APC), CNRS, Universite Paris 7 Denis Diderot,
10, rue Alice Domon et Leonie Duquet, F-75205 Paris Cedex 13, France
\thanks{UMR 7164 (CNRS, Universit\'e Paris VII, CEA, Observatoire de Paris)}
\and
Dublin Institute for Advanced Studies, 5 Merrion Square, Dublin 2,
Ireland
\and
Landessternwarte, Universit\"at Heidelberg, K\"onigstuhl, D 69117 Heidelberg, Germany
\and
Laboratoire de Physique Th\'eorique et Astroparticules, 
Universit\'e Montpellier 2, CNRS/IN2P3, CC 70, Place Eug\`ene Bataillon, F-34095
Montpellier Cedex 5, France
\and
Universit\"at Erlangen-N\"urnberg, Physikalisches Institut, Erwin-Rommel-Str. 1,
D 91058 Erlangen, Germany
% ...
}

\abstract
{}
{The X-ray--TeV connection and the evolution of the emitting particle 
population is studied in high-energy peaked BL Lac objects, by obtaining
spectral information in both bands on sub-hour timescales.
}
{Simultaneous observations with HESS, {\chandra} and the Bronberg optical observatory
were performed on the BL Lac object PKS 2155--304  in the night of July 29--30 2006,
when the source underwent a major $\gamma$-ray outburst 
during its high-activity state of Summer 2006. This event took place about 44 hours after the 
other major outburst of the night of July 27--28, which is known for its ultrafast variability.
An unprecedented 6 to 8 hours of simultaneous, uninterrupted coverage was achieved,
with spectra and light curves measured down to 7 and 2 minute timescales, respectively.
}
{The source exhibited one major flare along the night, at high energies.
The $\gamma$-ray flux reached a maximum of $\sim$11 times the Crab flux ($>$400 GeV), 
with rise/decay timescales of $\sim$1 hour, plus a few smaller-amplitude flares superimposed 
on the decaying phase. 
The emission in the X-ray and VHE $\gamma$-ray bands is strongly correlated,
with no evidence of lags. The spectra also evolve with similar patterns, and
are always soft (photon index $\Gamma$$>$2), indicating  no strong shift of the peaks 
in the spectral energy distribution towards higher energies. 
Only at the flare maximum is there evidence that the $\gamma$-ray peak is inside the observed
passband, at $\sim$400--600 GeV.
The VHE spectrum shows a curvature that is variable with time and stronger at higher fluxes.
The huge VHE variations ($\sim$22$\times$) are only accompanied by 
small-amplitude X-ray and optical variations (factor 2 and 15\% respectively).
The source has shown for the first time in a high-energy peaked BL Lac object
a large Compton dominance  (L$_{\rm C}$/L$_{\rm S}\sim$10) -- rapidly evolving --
and a \emph{cubic} relation between VHE and X-ray flux variations, during a decaying phase.
These results challenge the common scenarios for the TeV-blazar emission.
}
{}

\authorrunning{The HESS Collaboration}
\titlerunning{Simultaneous $\gamma$-ray/X-ray/optical observations of PKS\,2155--304 during an exceptional flare.}

\maketitle

\section{Introduction}
Among blazars, high-energy peaked BL Lac objects \citep[HBL;][]{giommipadovani94} 
are characterized by  the highest energy particles and  most violent acceleration processes. 
In the X-ray band, they have shown extreme spectral properties
\citep[see e.g.,][]{extreme} and variability \citep[e.g., Mkn 501,][]{pian98}.
At  very high energies (VHE, $\gtrsim$100 GeV),
doubling timescales as short as a few minutes and  flux variations of a factor of 10 
in less than one hour have been observed \citep{bigflare,MAGICMKN501}.
Their spectral energy distribution (SED) 
is dominated by two broad peaks, located at UV--X-ray frequencies and 
in the GeV--TeV band. 
Their overall properties and behaviour
have been most successfully -- though not exclusively -- explained 
as synchrotron and inverse Compton (IC) emission from a population of 
relativistic electrons 
\citep[see e.g.,][and references therein]{gg98,spada01,sik_mad,gcc2002,dafne04}, which upscatter
their own self-produced synchrotron photons 
\citep[synchrotron self-Compton, SSC;][]{konig,maraschi92,bloom96}
or external photons produced by different parts of the jet \citep{markos03,ggspinelayer}.
Target photons can also be provided by the accretion disk and broad line region 
\citep[BLR,]{sikoraec94,ggmadau96,dermer96} 
or by a dusty torus \citep{sikora94,ww95,sik_mevbl02}.
In general, all of these seed photons can contribute significantly to the production 
of the observed SED, according to their energy density in the comoving frame.
In HBL however, the lack of evidence of intense diffuse external fields 
(both directly from almost featureless optical-UV spectra and from 
TeV $\gamma$-rays transparency arguments)
has favoured the SSC model and external Compton process on photons from different 
parts of the jet as the most likely channels.

Providing two handles on the one electron distribution responsible for both emissions,
simultaneous observations in the X-ray and VHE bands 
represent both a powerful diagnostic tool and a very stringent 
testbed for the model itself  \citep{coppiahar99},
especially during large flares when the emission from a single region 
is expected to dominate the SED.
Alternatively, hadronic scenarios explain the $\gamma$-ray peak as being produced by 
ultra high energy protons \citep[$E\gtrsim10^{19}$eV; see e.g.][]{felix2000,anita01}.

Imaging atmospheric Cherenkov telescopes (IACT) provide a unique chance to study 
rapidly-variable sources at $\gamma$-ray energies, 
thanks to their large collecting area. 
Multiwavelength campaigns performed on a few very bright sources
(namely Mkn\,501, Mkn\,421 and 1ES\,1959+650)  
have shown that X-ray and VHE emission are  generally highly correlated
\citep[e.g.,][]{pian98,djannati99,hegra501spec,hegra501lc,sambruna2000,
501,1959,blazejowski05,giebels,fossati08}, 
down to sub-hour timescales with no evidence of significant lags 
\citep{maraschi99,fossati04,fossati08}.  
Moreover, the correlation seems to tighten  
when individual flares can be fully  sampled \citep{fossati08}.
These results provide strong support to the idea that 
both emissions during flares
are produced by a single electron population  (``one zone'' SSC scenario).

On the other hand, the same campaigns have also unveiled a more complex and puzzling 
behaviour, which represents a challenge to the SSC scenario. 
Two main problems have recently emerged:

1) the X-ray and VHE emissions do not always correlate 
\citep[e.g., in Mkn\,421;][]{blazejowski05}. 
In particular, VHE flares seem to occur also without any visible X-ray counterpart
(so-called ``orphan'' flares). The most striking example has been provided by 1ES\,1959+650 
during the high state of 2002, when a strong ($>4$ Crab) and rapid TeV flare 
(7 hours of doubling timescale) was not accompanied by detectable variations 
in the \rxte band \citep{1959}.
True orphan events are quite difficult to explain with a standard SSC scenario.
However, the generally sparse sampling does not allow the exclusion of
lagged counterparts \citep{blazejowski05}
or counterparts emerging in a different energy band \citep{1959}.

2) Mkn 421 exhibited a quadratic relation between VHE and X-ray flux variations
during both the rising \emph{and decaying} phases of a flare \citep{fossati04,fossati08}.
This is not expected if the source is in the Klein-Nishina (KN) regime.

A blazar is generally said to be ``in the Klein-Nishina regime'' when the observed
$\gamma$-ray emission is produced by TeV electrons  which do not upscatter their own 
self-produced synchrotron photons 
-- since inhibited by the smaller cross-section of the KN regime --
but  upscatter in the Thomson regime lower-energy photons 
produced by lower-energy electrons\footnote{Note that this case is different from
the condition where the observed $\gamma$-rays are indeed produced by IC-scatterings
occurring in the Klein-Nishina regime, or when the cooling itself is determined by 
Klein-Nishina losses, see e.g., \citet{moderski05}.}  
\citep[see e.g.,][]{tavecchio98}.
This condition changes the 
mapping of the synchrotron and Compton components,
so that the two peaks are not produced by electrons of the same energy.
In this situation, the VHE emission tends 
to track the X-ray synchrotron variations 
only linearly instead of quadratically,
although a flare that is achromatically extended  over a sufficiently wide energy range
can still yield 
a quadratic increase \citep{fossati08}. 
However, the energy dependence of both synchrotron  and IC cooling  ($\propto \gamma^2$)
prevents this relation 
during the decaying phase: since higher-energy electrons
cool faster than lower-energy electrons, 
they see a roughly constant seed-photon 
energy density, resulting in a mostly linear decrease.
A quadratic decrease could be achieved by imposing the strict Thomson condition,
but that seems to require extremely large beaming factors for Mkn 421 
\citep{katar05,fossati08}.

To investigate these issues, a multiwavelength study of single flares is essential.
Although many efforts have been made to achieve a good sampling, so far
the short  variability timescales have been difficult to study, 
because of the lack of sensitivity of the past-generation of IACTs. 
However, these are extremely interesting timescales:
the results of HEGRA on Mkn 421, for example, have already revealed 
an entire ``zoo'' of intra-night flares 
with different rise and decay times \citep{hegra421}, indicative
of a complex interplay between acceleration/injection and cooling processes
\citep[e.g.,][]{kirkmast99}

When the IACT array \hess became operational, 
a project was therefore developed
with specific ToO proposals, to investigate the fast variability timescales with
a whole night (6-8 hours) of continuous, simultaneous observations  during a bright $\gamma$-ray state.
To achieve this aim, {\chandra} was chosen because it is the \emph{only} X-ray satellite
capable of a full coverage of the entire \hess visibility window during most of the year,
and without the interruptions on sub-hour timescales  
which are typical of low-orbit satellites. 
The efforts paid off in July 2006, when the HBL  PKS\,2155--304 ($z$=$0.116$) 
became highly active at VHE, with a flux level a factor of $\sim$10 higher than
its typical quiescent flux of $\sim$4\e{-11} \cms above 200 GeV. 
PKS\,2155--304 is one of the brightest and 
most studied BL Lacs 
in the Southern Hemisphere, at every wavelength,  
and it can be detected by \hess on almost a nightly basis  since 2002
\citep[see][and references  therein]{hess2155,2155mwl}.

In the first days of activity, ToO observations were also triggered 
for other X-ray satellites such as \rxte and \swift, to sample
the source behaviour over several days and weeks.
Then, in the early hours of July 28 2006, 
a giant outburst occurred at VHE
($\sim$100$\times$ above the quiescent level),
with a peak flux of 15 Crab above 200 GeV (corresponding to 9.9 Crab above 400 GeV) 
and repeated flares with doubling timescales of few minutes \citep{bigflare}.
Unfortunately, this dramatic outburst occurred too early with respect to the
already-triggered but not-yet-started X-ray observations 
(which were acquired from the night after).
Therefore, the event of most exceptional variability remained
without  multiwavelength coverage.

Two days later however, on the night of July 29-30, 
the source underwent a second major VHE flare, this time in coincidence 
with our planned {\it Chandra}--\hess ToO campaign, and with the further coverage 
in the optical band provided by the Bronberg Observatory in South Africa. 
Snapshot observations of few ks were also taken with \rxte and \swift. 
This second outburst has reached even higher fluxes 
than the first one ($\sim$11 Crab above 400 GeV).
As a result, in this single night the most dense and sensitive X-ray/TeV campaign 
to date was obtained,  during one of the brightest states ever observed from an HBL at VHE.

This paper  focuses on the multiwavelength results of this exceptional night,
presenting new optical, X-ray and VHE data. 
The results of the whole VHE activity of PKS 2155--304 between July and October 2006
will be presented in forthcoming papers, together with the overall multiwavelength coverage.

Since this is a very rich and complex dataset, the
data have been divided into several subsets of different time windows,
to highlight specific aspects 
(e.g., different VHE thresholds, integration times, or X-ray coverage). 
These subsets and their rationale will be introduced in the relevant Sects.,
but a summary list with corresponding time windows
is given in Table \ref{tab:datasum}, for reference.
Throughout the paper, the following cosmological parameters are used:
$H_0\,=70$~km~s$^{-1}$~Mpc$^{-1}$, $\Omega_{\rm M}\,=$ 0.3, and $\Omega_{\rm \Lambda}\,=$ 0.7.
Conforming to the convention adopted in all previous \hess publications, 
unless otherwise indicated, all errors are given at the 1~$\sigma$ confidence level
for one parameter of interest ($\Delta\chi^2=1.0$). For simplicity,
in the text the values of MJD are given as \mjd$\equiv$\,MJD-53900.

\section{Observations and data reduction}
\subsection{HESS}
\hess is an array of four Imaging Atmospheric Cherenkov Telescopes located in the
Khomas Highlands of Namibia (23$^\circ$S, 15$^\circ$E, 1800 m a.s.l.). Each telescope has a
surface area of 107 m$^2$ and a total field of view of 5$^\circ$. A camera  consisting of 960
photo-multipliers is located at the focal length of 13 m. Each camera
images the dim Cherenkov flashes from  air-showers of VHE $\gamma$-rays in the
atmosphere, collected  by its mirrors (for more details about the layout
of the telescopes, see \citet{TELESCOPE1}).
The camera images are calibrated following the prescriptions in \citet{CALIB}.
The stereoscopic view of the air showers allows the reconstruction of the
direction of the primary $\gamma$-ray with an accuracy of $\approx$0.1$^\circ$ following 
method 1 of \citet{SHOWERRECO}.

The recorded signal in the field of view is dominated by the constant background from hadronic 
cosmic rays entering the atmosphere. 
Most of the hadronic background can be  identified from 
by the shape of the shower images and the arrival direction of the recorded showers.
The remaining hadronic background can be statistically removed by
estimating it from sky regions with no $\gamma$-ray signal. 
In the analysis shown here, \emph{loose cuts} and the reflected background
method were applied for background substraction \citep{CRAB}. 
Light curve and spectra were derived following the
standard \hess analysis also described in this reference.

\hess observed PKS 2155--304 throughout the  night of July 29-30.
A total of 15 runs (each 28 min long) were taken, all
passing the standard quality criteria specified in
\citet{CRAB}.
The total lifetime after dead-time corrections is 6.58 hours.
A $\gamma$-ray excess of 32\,612 events was detected with a
significance
of 254 $\sigma$ following Equation 17 in \citet{LIMA}. The
excess is point-like, taking into account the point spread function of HESS, with a
best fit position of $\alpha_{2000} = 21^h 58^m 52.6^s \pm 0.1^s_{stat} \pm 1.3^s_{sys}$, $\delta_{2000} =
-30^{\circ}13'29.8''\pm 1.9_{stat}'' \pm 20_{sys}''$, consistent with the position of PKS\,2155--304
\citep[$\alpha_{2000} = 21^h 58^m 52.0651^s , \delta_{2000} = -30^{\circ} 13' 32.118''$;][]{POSITION}.
Because of the long duration of the observations, 
the zenith angle of the source varied
strongly during the night, going from 53$^\circ$ at the beginning, 
to 8$^\circ$ at the middle, and back to 50$^\circ$ at the end of the night.
Observations at larger zenith angles imply
a higher energy threshold of the analysis.
For the observations discussed here the energy
threshold varies between 200 and 700 GeV for the applied energy
reconstruction \citep{CRAB}. This results in a tradeoff between energy
and time coverage for the analysis, as we discuss later.

\hess has a systematic uncertainty in the normalization of its
energy scale of $\approx$15\%. The main source of this systematic error are
uncertainties in the atmospheric conditions 
\citep[for a more detailed discussion see][]{CRAB}. During the night the atmospheric
conditions were stable. This can be verified to timescales
shorter than one minute in the overall trigger rate and background 
rates in regions off the source.
The differential energy spectrum $\Phi(E)$ of PKS 2155--304 at
VHE energies is generally steep,
with a photon index of about 3.4 in a power-law model $\Phi(E) = dN/dE
= \Phi_0 E^{- \Gamma}$
\citep{hess2155,2155mwl,bigflare}. The
systematic error in the energy scale of the detector therefore transforms into an error
of $\approx$40\% in the overall flux normalization. The systematic error
in the slope of differential energy spectra is $\approx$0.1 for the
photon index $\Gamma$
\citep[see][]{CRAB}.

\subsection{Chandra}
PKS 2155--304 was observed with {\chandra} \citep{chandra} for a total duration of 30 ks with the 
Low Energy Transmission Grating (LETG) spectrometer coupled with the ACIS-S detector 
(ObsId 6874; set-up with 1/8 subarray and 6 active chips, for a 0.7 s frametime).
%
Because of difficulties in placing our constrained ToO observation within the 
{\chandra} schedule,   the observation started  later than requested,
missing the first 1.7 hours of the \hess window. 
Unfortunately, the main flare occurred
in the first few hours of the \hess window. As a result, the rising part 
of the main VHE flare has no X-ray coverage. 

Data reduction was performed according to the standard CXC procedures,
using the CIAO software version 3.4 with the 
corresponding Calibration Database CALDB version 3.3.0 and HEASOFT v6.3.2.
Event files on timescales as short as  2 minutes  
were obtained using \verb+dmcopy+,  which propagates all deadtime corrections
correctly.
Grating spectra were then extracted with \verb+tgextract+
and their ancillary files were generated with \verb+fullgarf+ for each arm;
then added together to obtain the first-order spectrum.
The scientific analysis was completed mainly on the first-order spectrum, 
because of its higher S/N and photon statistics. 
A check was performed that the centroid of the source 
obtained from the zero-order image was indeed coincident 
with the coordinates of the source on the detector.
The response matrix was produced using  \verb+mkgrmf+ applied to the entire observation,
since no difference was found from files created in different epochs during the night.
The analysis was optimized and performed only on the continuum properties: the study
of the total grating spectrum at its highest resolution is beyond the scope of this paper
and is left to future publications.
The background and source spectra in each time-bin were then extracted with 
the tool \verb+tg_bkg+ for use in XSPEC.

The hundreds of spectra (one for each time bin,  down to 2-minute bins)
have been routinely fitted in XSPEC version 11.3.2, 
using a source model plus photoelectric absorption (\verb+wabs+), with the
equivalent hydrogen column density fixed at Galactic values
\citep[$N_{\rm H}=1.69\times10^{20}$ cm$^{-2}$;][]{dickey90}.
This is also the $N_{\rm H}$ value obtained from the best 
fit to the total exposure  with free absorption, to within 1 sigma.
The integrated flux and its error 
were calculated from the spectral fit using the specific  error routine in XSPEC. 
The error in the unabsorbed flux was then obtained from the percentage error
of the absorbed fit. The results of this procedure were later checked 
to be fully consistent  with  the values from the specific Tcl routine \verb+fluxerror.tcl+ 
recently provided by HEASARC with XSPEC v12, for calculating the error in the flux from single 
components of the model\footnote{See http://xspec.gsfc.nasa.gov/docs/xanadu/xspec/fluxerror.html}.
In the following, all X-ray fluxes are quoted as unabsorbed values.
The time analysis was also performed using the direct count rate and its error 
in the energy band of interest, for each time-bin, 
obtaining fully consistent results.  
The average count rate observed from PKS 2155--304 in the LETG 
is 8 cts/s, ranging between 12 and 6 cts/s. 
These count rates allow the spectra to be extracted down to 2--4 minute
timescales with typically 1000--2000 counts each.
For the observed flux, the grating spectra do not suffer from pile-up problems:
the total fraction of piled-up events estimated at the peak of the 
effective area (1--2 keV) and source peak flux is less than 5\%. 

One of the calibration issues with the ACIS instrument is the excess absorption
seen below 1 keV due to the build-up of contaminants on the optical blocking filter.
These contaminants (thought to be carbon compounds) cause a significant absorption 
in the 0.3--0.4 keV range, which is taken into account by the calibration but 
which also severely reduces the count rate in that range,
yielding very few counts during short exposures.  
Therefore, for the spectra extracted on 2 and 4 minute timescales,
the interval 0.3--0.4 keV was excluded from the analysis. 
At high energies, 
data were included in the fit up to the energy where positive net source counts were present.
As a result, short-timescale spectra were fitted in the range 0.23--0.3 and 0.4--6 keV,
while longer-exposure spectra could be fitted from 0.23 to 8--9 keV. 
The spectra were generally rebinned to have more than 30 counts per bin,
using different schemes.
A fixed coarse binning was used for all the spectra on short timescales 
($<7$ minutes).  Various checks have shown that, within the uncertainties, 
the obtained results are independent of the adopted rebinning.

The analysis of the total {\chandra} exposure shows  evidence of a slight residual
excess absorption in the 0.3--0.4 keV range, 
not yet fully accounted for by the calibration. The effect is small, and does not affect
the fit values significantly. Nevertheless, it was taken into account 
by simply adding an edge model at the carbon K edge, 0.31 keV,  with a fixed  value of $\tau=0.4$. 
These are the  best-fit parameters obtained from the fit of the total {\chandra} spectrum
(see Sect. \ref{sect:xspectra}).

\subsection{RossiXTE}
As part of the ToO campaign for a daily monitoring, {\it RossiXTE} \citep{pca} pointed 
PKS 2155--304 twice during the night of July 29--30, for a total exposure of $\sim$2.6 ks (44 minutes).
Fig. \ref{fluxgo} shows the epochs of the \rxte windows with respect 
to the overall \hess and {\chandra} light curves, together with the time-windows 
of the  \swift~ observation. The latter is analyzed and discussed in \cite{foschini2155}.
Since the \swift~ passband mainly overlays 
the {\chandra} band,
we focus only on the \rxte spectrum,  to 
extend the X-ray spectrum in the hard band.
The data reduction and analysis were performed using FTOOLS v6.3.2 
with the standard procedures and filtering criteria recommended by 
the \rxte Guest Observer Facility after September 2007
\footnote{See http://www.universe.nasa.gov/xrays/programs/rxte/pca/doc/bkg/bkg-2007-saa/}.
For a more accurate spectral determination, only the PCU2 data were considered.
The average net count rate from the source is measured at 6 cts/s/pcu, 
in the 3--20 keV band.

The \rxte spectrum was then fitted in XSPEC together with the {\chandra} spectrum 
extracted from the same time window, obtaining 
a spectral measurement over two decades in energy (0.2--20 keV).
Without adjustments, the two spectra have very similar normalizations,
indicative of a very good inter-calibration between the two instruments.
To obtain the most accurate spectral determination, the \rxte/{\chandra}
normalization was fixed at the value  measured by fitting the same power-law model
in the overlapping energy range (3--7 keV), namely 1.08 (see Sect. \ref{sect:xspectra}).

\subsection{Optical data}
Optical observations in the V filter were performed using the
32 cm Schmidt-Cassegrain telescope at Bronberg Observatory, Pretoria,
South Africa 
\citep[see][for details about observations of variable sources with this telescope]{imada}.
After de-biasing and flat fielding, data were analyzed using relative aperture photometry
with a K-type star of similar magnitude (12.6).
Frames were taken every 30 s and then smoothed over 6 successive data points
to calculate the mean value in each bin.  Each point of the optical
light curve therefore has a time duration of $\sim$180s.

The comparison with the reference star  shows that the rise in optical
intensity is highly significant.  The K-star light curve is constant with an
intrinsic scatter in the datapoints -- using the same smoothing procedure -- of
$\lesssim$0.02 mag.
The errors were determined from the variance of each 6 successive data points
for both the source and reference star. No time variability in the intensity of
the reference star is seen.
The optical fluxes were corrected for Galactic extinction  with
A$_{\rm V}$=0.071 mag. 

\begin{center} 
\begin{figure*}[ht]
\centering
   \includegraphics[width=19cm]{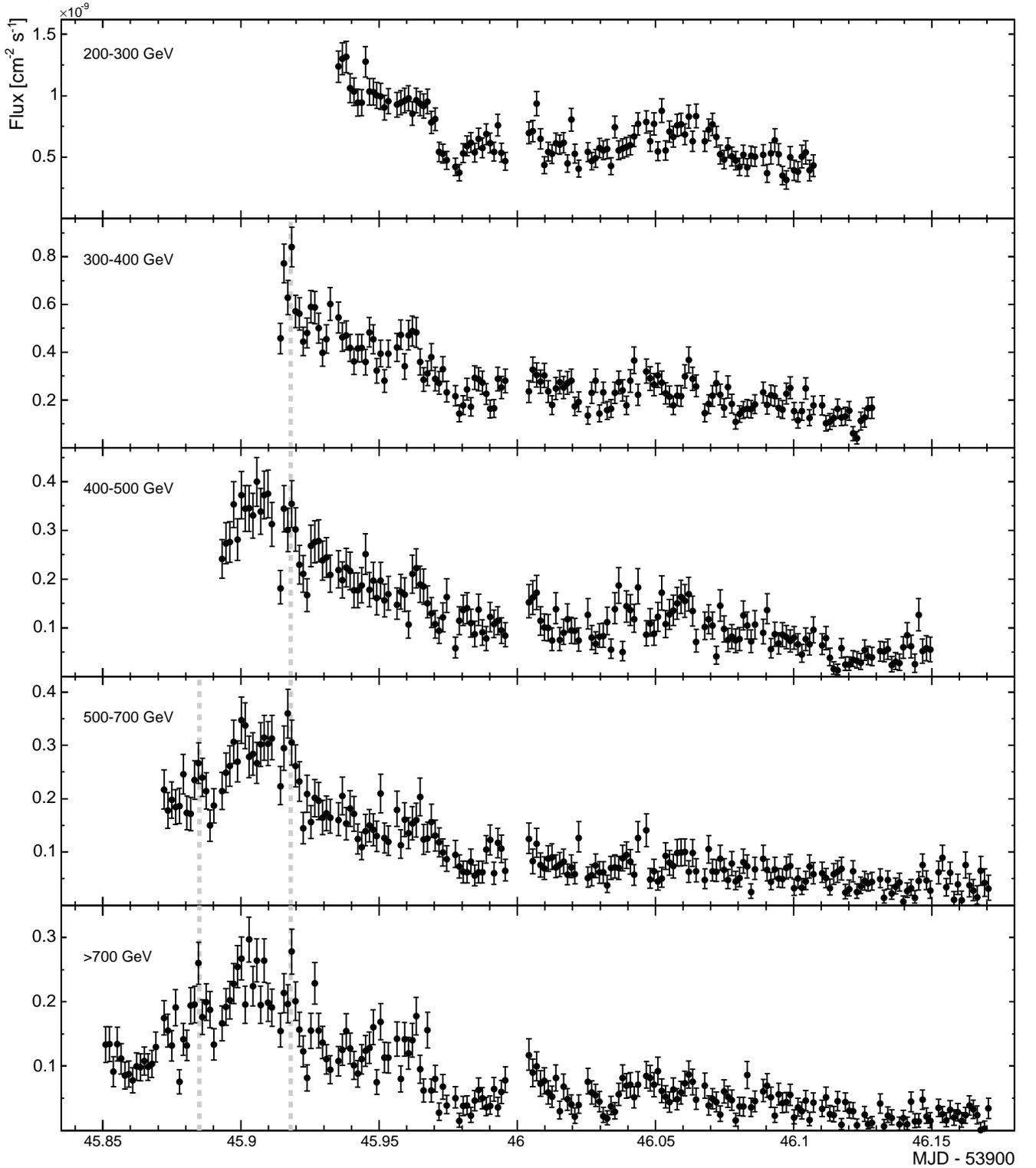}
     \caption{VHE fluxes integrated in different energy bands, as a function of time, 
     in time bins of two minutes. The time windows corresponding to the 
     different energy thresholds are given in Table \ref{tab:datasum} 
     (labelled accordingly, from top to bottom: \texttt{T200} to \texttt{T700}).
     The dotted lines mark the positions of rapidly varying events 
     on-top of the main flare (see text). }
     \label{fig:lcs2min}
\end{figure*}
\end{center}

\section{Temporal analysis}
\subsection{TeV light curves}
The measured VHE light curve in two-minute time bins is shown 
in Fig. \ref{fig:lcs2min}, divided into five different energy bands. 
The time coverage increases with threshold energy,
due to the aforementioned  zenith-angle effect. A full coverage of the
observation is thus reached for an energy threshold of 700 GeV,
while light curves (and spectra) down to $\sim$300 GeV are obtained
only for the central five hours of the observing window (see Table \ref{tab:datasum}).
The different time windows have therefore been labelled according
to the respective energy thresholds (from \texttt{T200} to \texttt{T700}), for reference.

\begin{table*}
  \begin{minipage}[t]{2.0\columnwidth}
    \caption{Summary of the subsets of VHE data used in this
  paper (\mjd=MJD-53\,900). 
  }
\label{tab:datasum}
\centering
\begin{tabular}{lccccccl}
\hline
\hline
\vspace*{-3mm}\\
Label\footnote{The T-label corresponds to the VHE energy threshold of that dataset.}     
&   \mjd start &  \mjd end  &   Duration & En. thres.  & Fig.\footnote{Figure where the 
time intervals are indicated and the Sect. where they are first introduced.} 
& Sect.$^b$ & notes \\
         &               &            &    [hrs]   &    [GeV]       &         &    &  \\
\hline
\texttt{T200}  &  45.934618  &  46.107844  & 4.16  & 200  &\ref{fig:lcs2min}& \S 3.1 & \\
\texttt{T300}  &  45.913536  &  46.129121  & 5.17  & 300  &\ref{fig:lcs2min}& \S 3.1 &\\
\texttt{T300-X}&  45.923252  &  46.129121  & 4.94  & 300  &\ref{fig:xg300}  & \S 3.3 & {\scriptsize \texttt{T300} with X-ray coverage}\\
\texttt{T400}  &  45.892483  &  46.150125  & 6.18  & 400  &\ref{fig:lcs2min}& \S 3.1& \\
\texttt{T500}  &  45.871450  & 46.171258   & 7.20  & 500  &\ref{fig:lcs2min}& \S 3.1& \\
\texttt{T700}  &  45.850393  &  46.171258  & 7.70  & 700  &\ref{fig:lcs2min}& \S 3.1& \\
            &             &             &       &      &            &  & \\
\texttt{T400-Peak}  &  45.896643  & 45.920474	& 0.57  & 400  &\ref{fig:fluxindex500}& \S 4.1& {\scriptsize Spectrum at the peak of VHE flare} \\
\texttt{T300-High}  &  45.913530  &   45.970312 & 1.36  & 300  &\ref{fig:fluxindex300}& \S 4.1& {\scriptsize Average high-state spectrum inside \texttt{T300}.} \\
\texttt{T300-Low}   &  46.013252  &   46.129166 & 2.78  & 300  &\ref{fig:fluxindex300}& \S 4.1& {\scriptsize Average low-state spectrum inside \texttt{T300}.}\\
            &             &             &       &      &            &  & \\
\texttt{T300-Xmax} &  45.922130  &  45.944699  & 0.54  & 300  &\ref{fluxgo}& \S 4.2& {\scriptsize Highest simultaneous X-ray/VHE state.}  \\
\texttt{T400-Xmin} &  46.100000  &  46.150000  & 1.20  & 400  &\ref{fluxgo}& \S 4.2 & {\scriptsize Lowest simultaneous X-ray/VHE state.}\\
\texttt{T300-RXTE} &  46.016846  &  46.033328  &       &      &  	  &  & \\
                   &  46.084624  &  46.098698  & 0.73  & 300  &\ref{fluxgo}& \S 4.1 &{\scriptsize Sum of the 2 intervals with RXTE coverage.} \\
\vspace*{-3mm} \\
\hline
\end{tabular}
\end{minipage}
\end{table*}

The source underwent a major flare in the first hours of observation,
reaching a peak flux of $\sim$9.9\e{-10} \cms ($>$400 GeV) at \mjd 45.90, 
corresponding to $\sim$11 times the Crab Nebula flux \citep{CRAB} 
above the same threshold. This peak flux is about 20\% higher than the peak 
flux measured during the night of July 27--28, above the same threshold
($\sim$9 Crab, $\sim$8\e{-10} \cms $>$400 GeV).
The fluxes measured in these two nights (July 27--28 and 29--30) 
are the highest ever observed at VHE from PKS 2155--304. 
The total amplitude of flux variation during this night 
was about a factor of $\sim$20, above 400 GeV
(\texttt{T400} covers both the highest and lowest flux epochs),
similar to the flux variation observed in the night of July 27--28 
($\sim$23$\times$).

The main flare seems to occur with similar rise and decay timescales,
of the order of 1 hour  
\citep[half-to-maximum amplitudes, measured using a ``generalized Gaussian'' function as in][]{bigflare}.
After the peak, the VHE flux decreased overall during the night reaching 
its minimum around \mjd 46.12,  
but with two other smaller-amplitude flares superimposed:  
a short burst around \mjd 45.96 of duration $\sim$20 minutes, and a
longer flare or plateau between \mjd 46.0 and 46.1, with a duration of 2--3 hours.

In addition, two further sub-flares are evident in all covered energy bands, 
around \mjd 45.885 and \mjd 45.920 (dotted lines in Fig. \ref{fig:lcs2min}).
These structures  have a duration of $\sim$10 min, similar to the flares
of the night of July 27--28. 
Although there are hints of even shorter variability (few minutes),
the significance is limited.

\subsection{Comparison with X-ray and optical light curves}
\begin{figure}[t]
\centering
   \includegraphics[width=9cm]{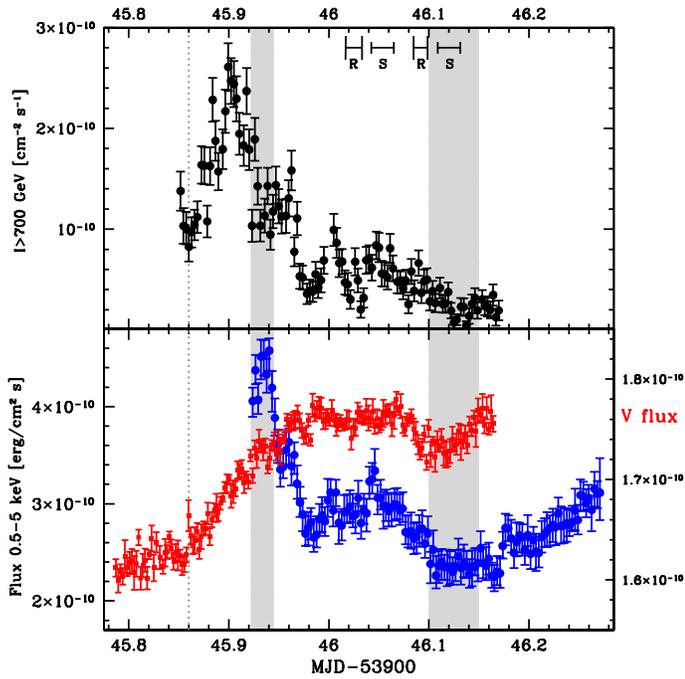}
     \caption{Overall light curves of PKS 2155--304 in the night of July 29-30 2006,
   as seen by \hess (\texttt{T700}, upper panel), {\chandra} (lower panel, blue circles), and 
   the Bronberg Observatory (optical V band, red squares). Time bins of 4 minutes (3 for the V band). 
   The segments on the upper x-axis also show the two intervals corresponding to the 
   \rxte exposure (R label), and the two intervals of the \swift~ pointing (S label) reported in
   \citet{foschini2155}.  \emph{The vertical scales differ} in each panel,  
   and have been adjusted to highlight the specific variability patterns.   
   Lower panel: the left axis gives the integrated 
   0.5-5 keV flux, the right axis gives the V-band \nufnu flux  at the effective frequency 
   5500 \AA;  both are in units of erg cm$^{-2}$ s$^{-1}$. The vertical line marks a
   visual reference time  for the start of both the optical and VHE flares.
   The shaded bands mark the time interval where the \texttt{T300-Xmax} and 
   \texttt{T400-Xmin} dataset are extracted
   (highest and lowest X-ray/VHE state; see Table 1). 
   }
     \label{fluxgo}
\end{figure}

\begin{figure}[t]
\centering
\vspace*{0.3cm}
\includegraphics[width=9cm]{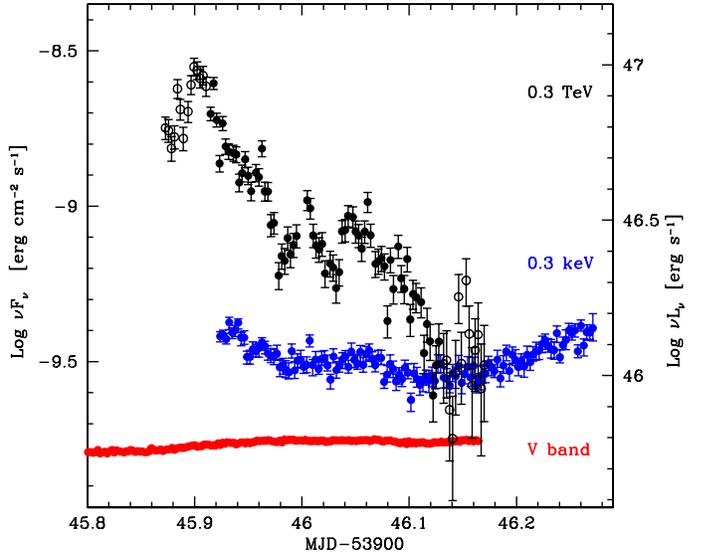}
\caption{Light curves of the \nufnu flux at 0.3 TeV, 0.3 keV, and 5500\AA~
in 4-minute bins, plotted on the \emph{same flux scale}. 
The right axis reports the corresponding luminosity scale.
Note the remarkable difference of the amplitudes and the dramatic evolution in 
the VHE/X-ray flux  ratio. 
\hess data (black): filled circles are fluxes calculated from the 
\texttt{T300} light curve (integrated above threshold);
empty circles  from the \texttt{T500} light curve (see text).
The 0.3 TeV fluxes are corrected for intergalactic $\gamma$-$\gamma$ absorption
with a low-density EBL model, 
while both X-ray and optical fluxes are corrected for Galactic extinction (A$_{\rm V}$=0.071). }  
     \label{nufnu}
\end{figure}

The combined VHE, X-ray and optical light curves are shown in 
Fig. \ref{fluxgo} and \ref{nufnu}.  Significant 
flaring activity is 
visible in all three bands, but with different amplitudes.
To emphasize the specific variability patterns,
the vertical scales in Fig. \ref{fluxgo}  were adjusted 
differently for each band. 
Fig. \ref{nufnu} shows instead the light curves on the same flux scale, 
but with a \nufnu representation. They correspond to slices of the SED at
the three energies of 0.3 TeV, 0.3 keV, and 2.25 eV (i.e. 5500\AA).  
In this way, it is possible to highlight the overall changes and time evolution 
SED-wise.

The 0.3 TeV fluxes were calculated from the integrated $>$300 GeV light curve 
(\texttt{T300} window) using the average power-law spectrum measured in the respective epochs 
(namely, the \texttt{T300-High} spectral index in the high state,
and the \texttt{T300-Low} index elsewhere, see Sect. 4.1).
The same procedure was used to calculate the 0.3 TeV fluxes  
from the $>$500 GeV light curve (\texttt{T500} window), 
during the epoch not covered by \texttt{T300}  (empty circles in Fig. \ref{nufnu}).
A comparison with the results of the $>$300 GeV light curve 
in the overlapping window 
shows that the extrapolation from 500 GeV does not introduce differences 
of more than $\sim$2\%.
%
The 0.3 TeV fluxes were then  corrected for the absorption effect 
caused by $\gamma$-$\gamma$ interactions
with the Extragalactic Background Light (EBL), using
the model by \cite{franceschini} 
(discussed in Sect. 4.1.1).
This corresponds to a low density of the EBL, close to  the
lower limits obtained by galaxy counts. 
The plotted fluxes therefore can be considered as lower limits to the 
intrinsic VHE emission of the source.
%
%
The X-ray fluxes at 0.3 keV  were calculated with the same procedure, 
using the power-law spectrum fitted in each of the individual bins. 
Anticipating  the result that both the VHE and X-ray spectra are steep 
($\Gamma$$>$2; see Sect. 4), the plotted \nufnu fluxes 
-- at the low-energy end of the respective passbands -- 
provide an estimate of the emission closer to the respective SED peaks
than the \nufnu fluxes in the hard band.

Several remarkable features can be noted.
The first is the huge difference in amplitude between the variations in the three energy bands.
In few hours, the VHE flux changed by more than an order of magnitude,
whereas the X-ray flux varied by only a factor $\sim$2 and the optical flux by 
less than 15\%
(the contribution of the host galaxy is negligible, see Sect. 6.3).
The source thus shows a dramatic increase in variability with photon energy.

Secondly, the VHE emission dominates the energy output by far in
the three bands. 
In Fig. \ref{nufnu}, the comparison
between the 0.3 TeV and 0.3 keV fluxes 
shows the evolution of the Compton dominance of the source,
i.e. the ratio of the IC $\gamma$-ray luminosity to the synchrotron power ($L_{C}/L_{S}$). 
The $\gamma$-ray luminosity dominates the synchrotron luminosity by a factor 
$\sim$8 close to the flare maximum, 
evolving rapidly towards comparable levels at the end of the night.
This is the first time that such high and rapidly variable Compton dominance 
is observed in an HBL, irrespective of the choice of  EBL density.
The swiftness of the changes in the $\gamma$-ray emission and SED properties 
also underlines the danger in modeling  X-ray and VHE data
taken even only a few hours apart, during such events. 
As shown in Fig. \ref{fluxgo},  both \rxte and \swift~ observations 
occurred when the Compton dominance had already decreased significantly.

Thirdly, despite this difference in amplitude, the X-ray and VHE 
light curves are strongly correlated, with the X-ray emission following closely the same pattern 
traced by the VHE light curve (see next Sect.).  
The optical light curve, instead,  does not correlate on short timescales with 
the other two bands. However, there is a rise of $\sim$15\% in flux, which appears
to begin at the same time as the main VHE flare.
A conservative estimate of the chance probability of coincidence
--considering only the data of this night-- is of the order of a few percent. 
We discuss in Sect. 8.2 the possible implications if this simultaneity is genuine.


\subsection{Inter-band time lags}
The degree of correlation and possible time lags between different
light curves have been quantified by means of cross-correlation
functions.
The correlation analysis was performed between X-ray and $\gamma$-ray
light curves and between
hard and soft energy bands within each
passband.
As main tool it was used 
the discrete correlation function (DCF) from \citet{DCF}. 
The DCF is especially suited to unevenly spaced data, such as the VHE
light curves, which have
gaps of a few minutes every 28 minutes between the stop and start of consecutive runs.  
The time lags between light curves are determined to be 
the maximum of a Gaussian plus linear function
fitted to the central peak of the DCF.  This procedure is robust against spurious peaks 
at zero time-tag caused by correlated errors \citep{DCF}.
The error in the measured time lag is determined by simulations.
Ten thousand light curves were generated by varying 
each measured point within its errors. 
The entire correlation procedure was repeated for each of these simulations,
resulting in a cross correlation peak distribution (CCPD). 
The RMS of the CCPD then provides an estimate of the statistical error
in the measured time lag  \citep{MAOZ,PETERSON}.
The time binning of the light curves and the DCF introduces an additional systematic
error. The latter was estimated to be 30~s by injecting various artificial
time shifts into the original VHE photon list, smaller than the
duration of the time bins, and by measuring the relative
shift of the CCPD.

\begin{figure}
\centering
   \includegraphics[width=9cm]{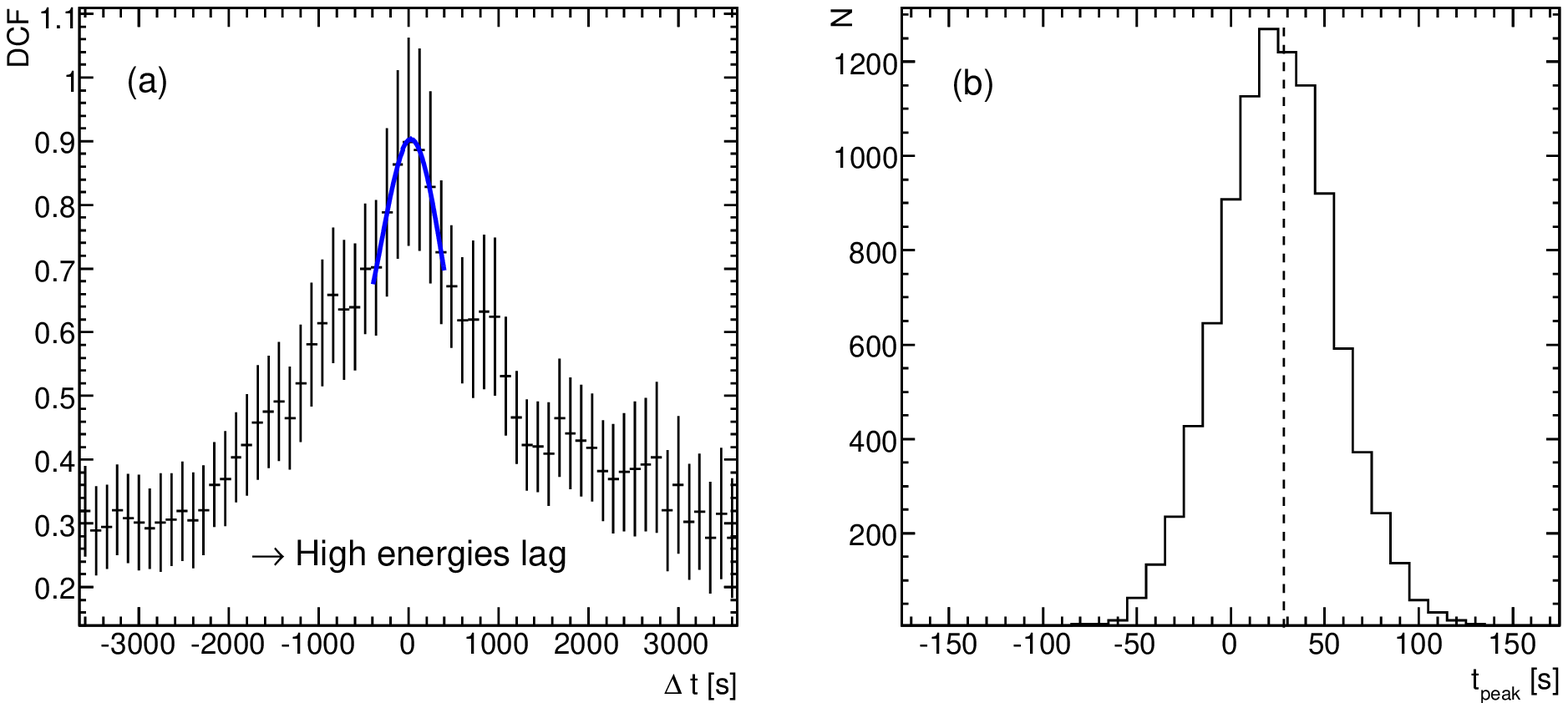}
   \includegraphics[width=9cm]{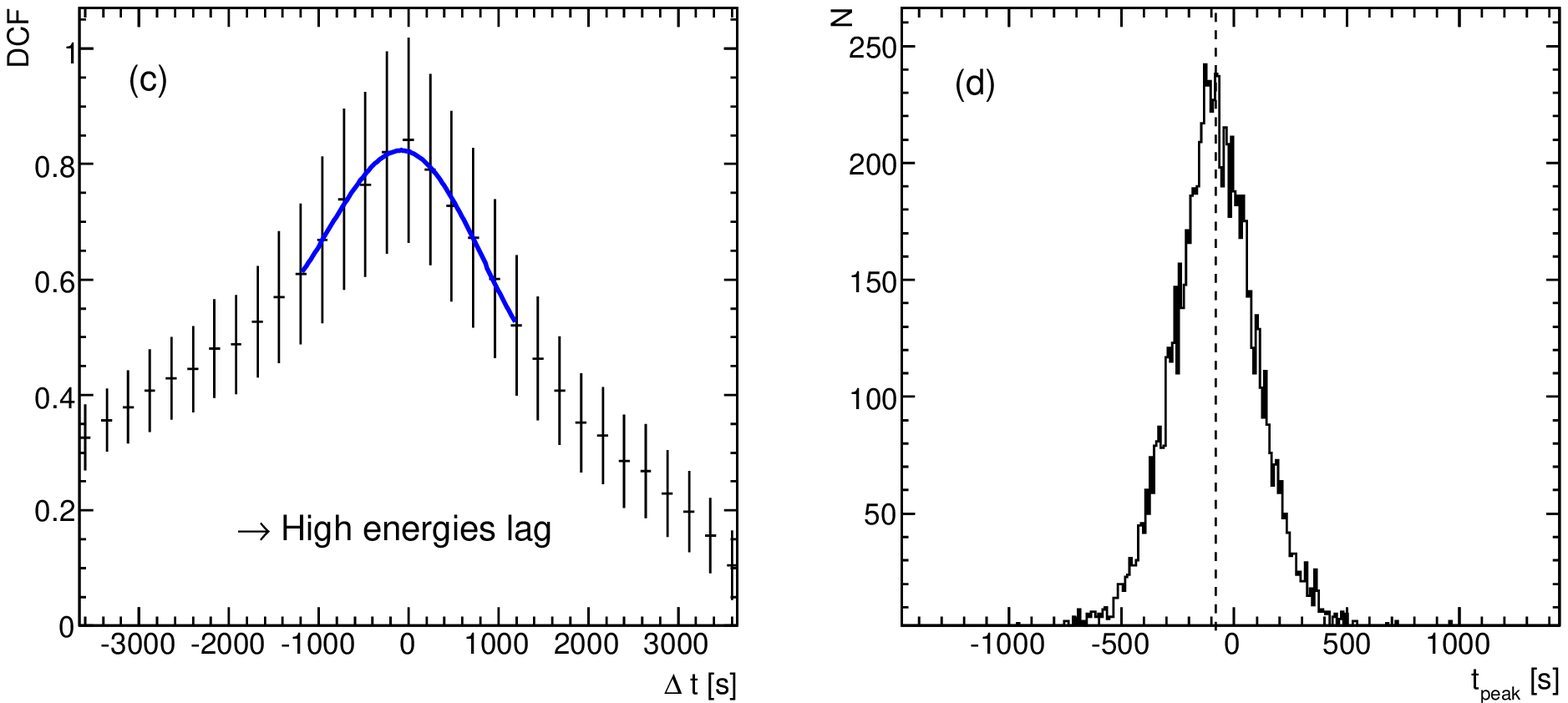}
     \caption{Cross-correlation analysis of the hard versus soft bands at VHE (upper panel)
     and at X-ray energies (lower panel).
     Upper panels: 
     (a) DCF of the 300--700 GeV and the $>$700 VHE band. The line around the peak shows the
     best fit Gaussian plus linear  function, with a maximum at 28~s.
     (b) Corresponding cross-correlation peak distribution (CCPD) of 10\,000 simulated
     light curves. The RMS of the distributions is 30~s. The dotted
     line marks the position of the maximum in (a).
     Lower panels:
     (c) DCF of the 0.2--1.0 keV and 2.0--6.0 keV X-ray
     band. The line around the peak shows the best fit Gaussian plus linear function, with a maximum 
     at -82~s.
     (d) Corresponding CCPD of 10\,000 simulated light curves. The RMS of
     the distributions is 202~s. The dotted
     line marks the position of the maximum in (c).}
     \label{fig:corrtev}
\end{figure}

\begin{figure}[t]
\centering
   \includegraphics[width=9cm]{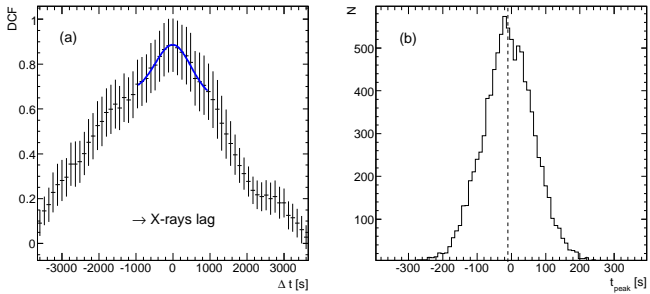}
     \caption{Cross-correlation between the X-ray and VHE light curves:
      (a) DCF of the $>$300 GeV light curve and the LETG 0.2--6 keV
      band.The blue line shows the best fit Gaussian plus linear
      function, with a maximum at -10~s.
      (b) 
      Corresponding cross-correlation peak distribution of 10\,000 simulated light curves. The RMS of
      the distributions is 76~s. The dotted
      line marks the position of the maximum in (a).
     }
     \label{fig:corrchandra}
\end{figure}

\begin{figure}[t]
\centering
   \includegraphics[angle=-90,width=9cm]{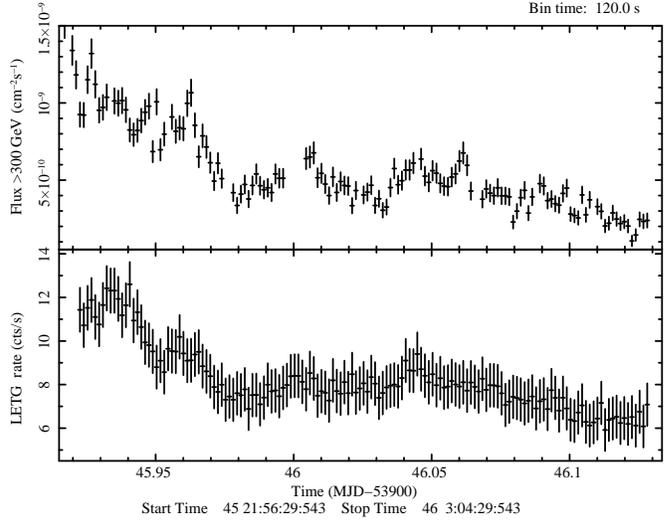}
      \caption{\hess and Chandra light curves in the simultaneous time windows   
     corresponding to a 300 GeV threshold (\texttt{T300-X} in Table 1). Two-minute time bins. 
     Upper panel: integral fluxes above 300 GeV. Lower panel: 1st-order LETG count rate
     in the 0.2--6 keV band.
      }
     \label{fig:xg300}
\end{figure}

At VHE, the analysis was performed on the simultaneous light curves between 300--700
GeV and above 700 GeV in two-minute time bins (Fig. \ref{fig:lcs2min}). 
This choice yields a
good compromise between event statistics, time coverage, and a maximum
energy difference between the bands.  The resulting time lag between
the higher and lower energy band is
(28~$\pm$~30$_{\rm stat}$~$\pm$~30$_{\rm sys}$)~s (see Fig. \ref{fig:corrtev}). 
This time lag does not differ significantly from zero, 
as for the flare of two nights before \citep{quantumg},
and we derive a 95\% confidence upper limit of 129~s. 
This value was calculated by assuming a Gaussian
probability distribution around the measured time lag. The width of the distribution was set
to be the linear sum of the statistical and systematic error, to
be conservative. Afterwards
symmetric intervals around zero were integrated, until a 95\%
containment was achieved. 

In the X-ray band, an analogous procedure was applied. 
The total {\chandra} light curve  was divided into a soft (0.2--1.0 keV) and
hard (2.0--6.0 keV) band.  Because of the larger errors, in this case
4-minute time bins were used. The measured time lag is
(-82~$\pm$~202$_{\rm stat}$~$\pm$~30$_{\rm sys}$ )~s.
This value again does not differ significantly from zero,
resulting in a 95\% upper limit of 482~s.

The cross-correlation analysis between the X-ray and  $\gamma$-ray emission was
performed on the simultaneous  light curves with two-minute time bins
shown in Fig. \ref{fig:xg300} (in the \texttt{T300-X} time window).
The resulting cross-correlation is shown in Fig. \ref{fig:corrchandra}.
The two light curves overall are  highly correlated, with a maximal correlation of
DCF$_{\rm max}\approx 0.9$, and no significant lag is found. 
The time lag of the X-rays with respect to the $\gamma$-ray is
(-10~$\pm$~76$_{\rm stat}$~$\pm$~30$_{\rm sys}$)~s, yielding
a 95\% confidence upper limit to a time lag of 208~s.
To test whether this result was caused by an averaging of lags with opposite signs, the
correlation  analysis was also performed on sub-intervals, 
namely in the interval around the first small flare at \mjd 45.96 (\mjd 45.94--46.0) 
and after \mjd 46.0.  
The two emissions are again highly correlated
(DCF$_{\rm max}\approx0.9$), with no evidence of time lags. 
The strong  correlation is determined not merely by the overall decaying trend 
of both light curves, but also by their specific patterns:  
a DCF max value of $\approx0.7-0.8$ is still obtained after whitening the light curves  
by removal of either a linear or quadratic trend.
On the shortest timescales ($<$4--8 minutes), however, the VHE light curve
shows few small flares apparently not mirrored in the X-ray band
(see e.g. \mjd 45.925 and 46.060 in Fig. \ref{fig:xg300}).
Although this might indicate a more complex correlation on the fastest timescales, 
at present no firm conclusions can be drawn, since 
the significance of these structures is low.

The correlation analysis was also performed using the task CROSSCOR
of the Xronos 5.21 package, which measures the correlation function (CCF)  with a
direct Fourier algorithm. This algorithm requires
a continuous light curve without interruptions, therefore
the few gaps were filled with the running mean value calculated over the 8 
closest bins \citep[e.g.,][]{ravasio421}.
The results are fully consistent with the DCF analysis,
indicating that the small gaps in the VHE light curves -- the X-ray light curve is
continuous -- do not introduce significant distortions  for such
well sampled data. 

The cross-correlation analysis between VHE and X-ray light curves
was limited to the strictly simultaneous window, to avoid artifacts in the 
lag determination due to the different timespans and the light curves characteristics.
Because both light curves have each one main flaring feature, the cross-correlation 
performed on different intervals tends simply to match the maxima of the two emissions
in those intervals, irrespective of the smaller amplitude patterns. 
This would yield an artificial, window-dependent  ``lag" with typically 
lower correlation values (as is the case here, with a timespan of $\sim$3200 s
between the maxima of the overall VHE and X-ray light curves 
and lower DCF/CCF values $\sim$0.7).

\begin{figure*}[t]
\centering
   \includegraphics[width=9cm]{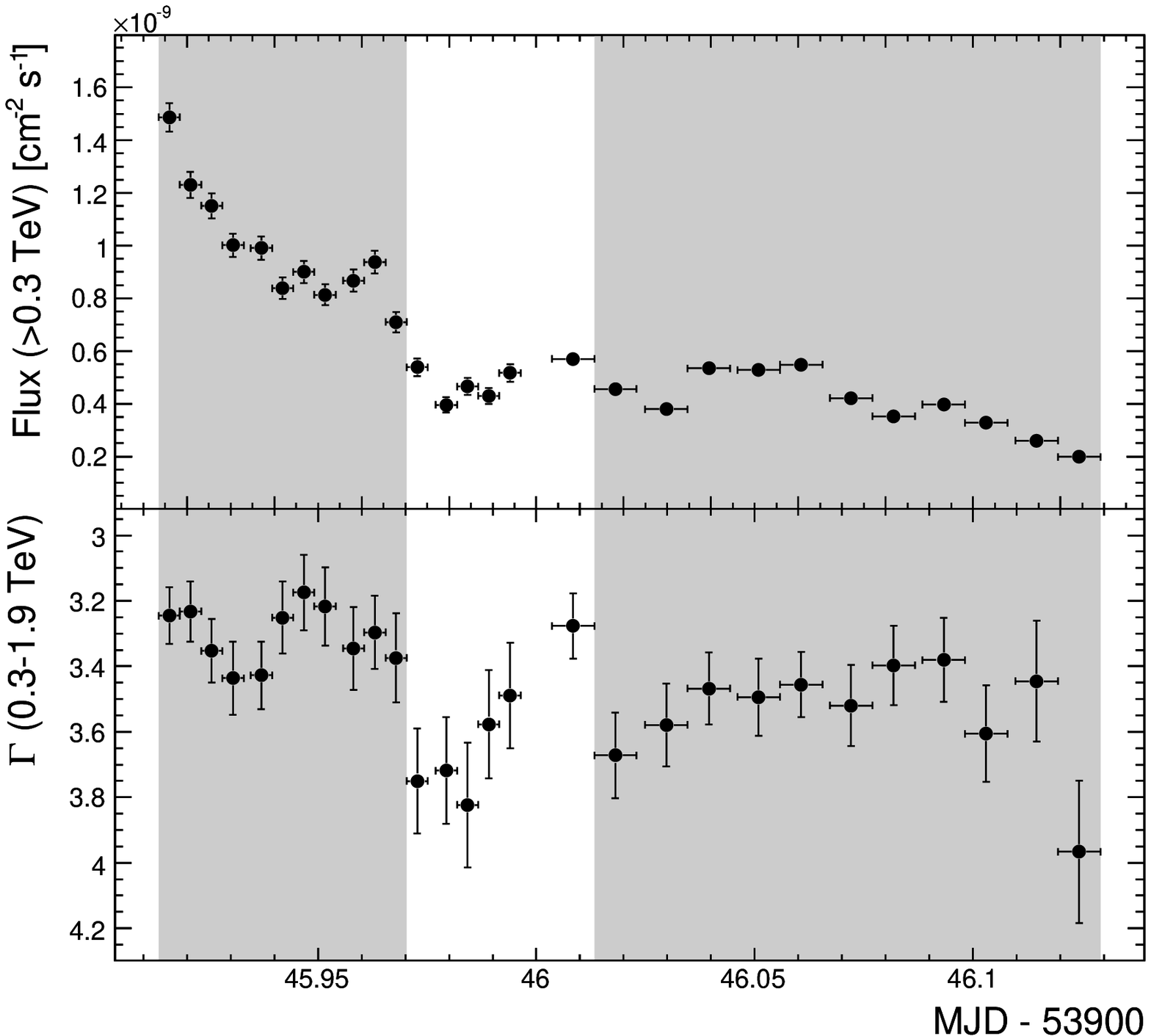}
   \includegraphics[width=8cm]{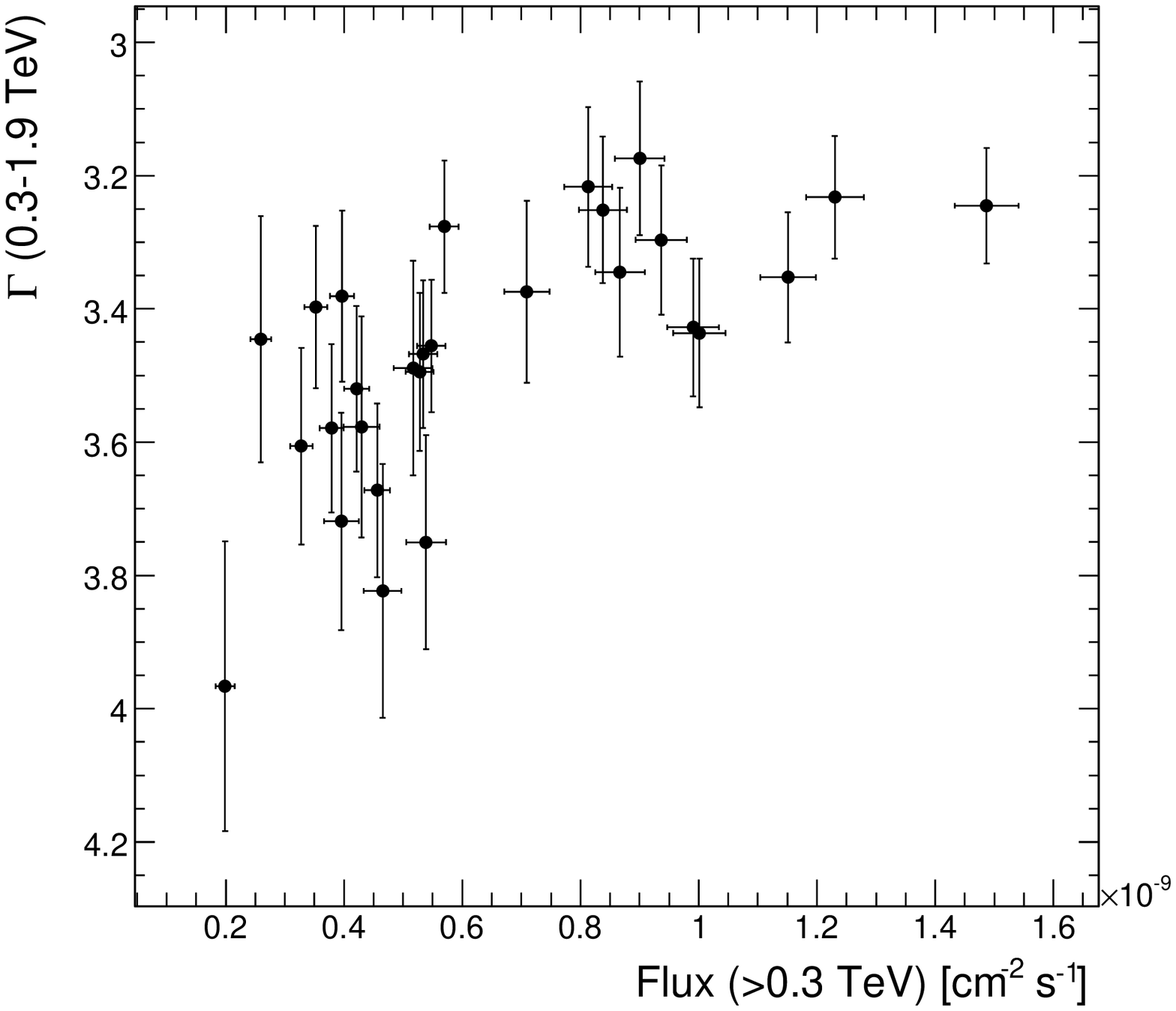}
     \caption{Left panel: integral flux $>$300 GeV and photon index
     as a function of time (\texttt{T300} dataset). Horizontal error bars show the time
     interval of each bin, going from 7 to 14 minutes before and after \mjd 46.0.
     The shaded zones mark the two time-intervals corresponding to the average 
     high and low-state spectra fitted in Table \ref{tab:specs300} 
     (\texttt{T300-High} and \texttt{T300-Low}, respectively). 
     Right panel: photon index as a function of the integral flux.} 
     \label{fig:fluxindex300}
\end{figure*}

\begin{figure*}[t]
\centering
   \includegraphics[width=9cm]{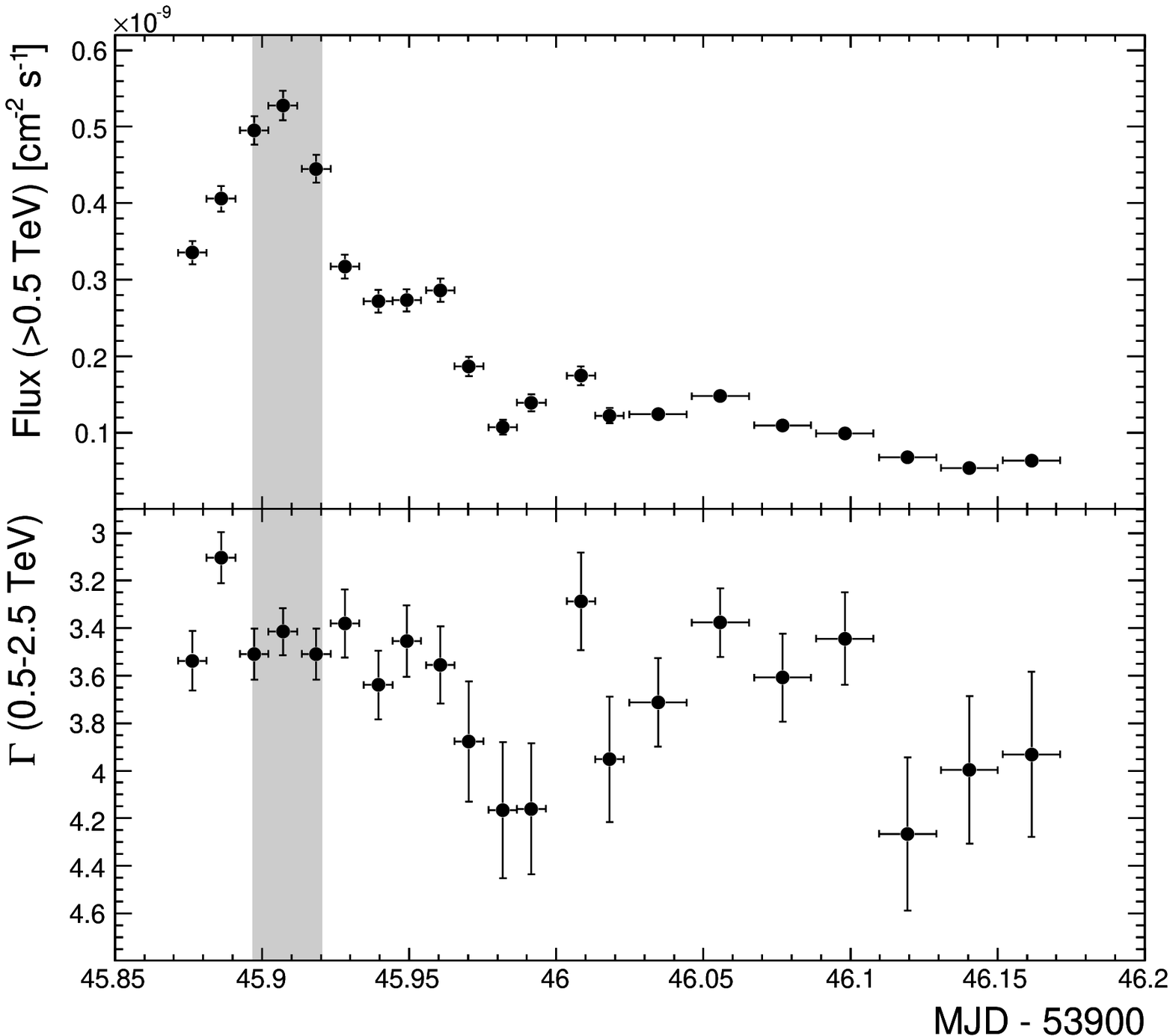}
   \includegraphics[width=8cm]{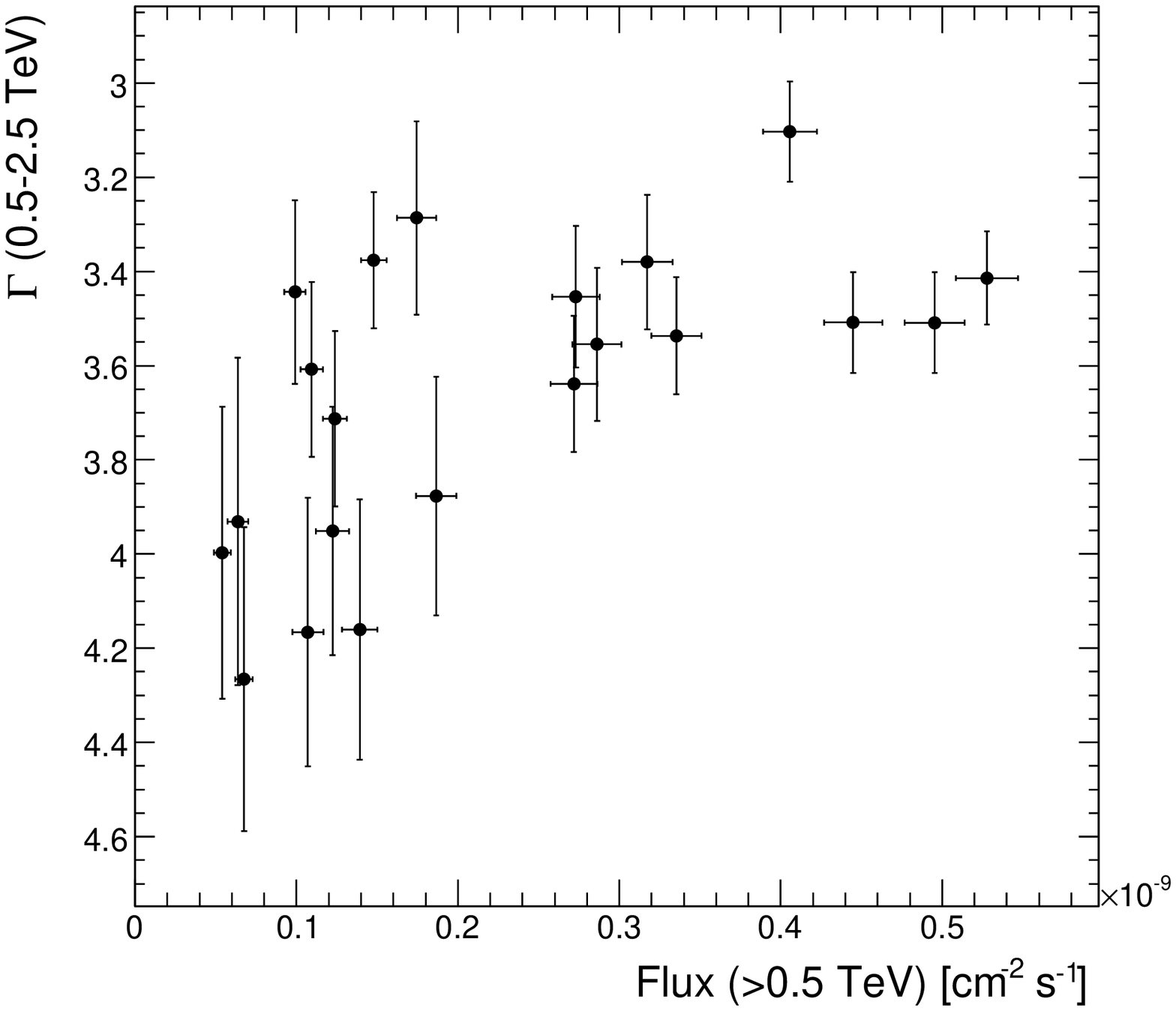}
     \caption{Left panel: integral flux $>$500 GeV and photon index
     as a function of time (\texttt{T500} dataset). Horizontal error bars show the time
     interval of each bin, going from 14 min. at the beginning to
     28 min. towards the end of the night. The shaded area shows the time window
     where the $\gamma$-ray peak spectrum (\texttt{T400-Peak} dataset 
     in Tables \ref{tab:datasum},\ref{tab:specs300})
     has been extracted from.
     Right panel: photon index as a function of the integral flux.} 
     \label{fig:fluxindex500}
\end{figure*}

\section{Time-resolved spectral analysis}
\subsection{VHE spectra}
A search for spectral variations in the  VHE data was performed by fitting a
power-law spectrum to a fixed energy range in fixed time bins. The
unprecedented statistics of this dataset allow the sampling 
in 7 to 14-minute bins in the \texttt{T300} time window (Table 1).
On these short integration times, the power-law function gives a
statistically good description of the data. 
The results are shown in Fig. \ref{fig:fluxindex300}. 
The spectrum generally hardens with increasing flux. 
The fit to a constant photon index
results in a $\chi^{2}$ probability of only 1.6\%.
These spectra are also used for comparison with 
the X-ray spectra extracted in exactly  the same time bins,
and discussed in  Sect. \ref{sect:xtev}.

\begin{figure*}[t]
\centering
   \includegraphics[width=18cm]{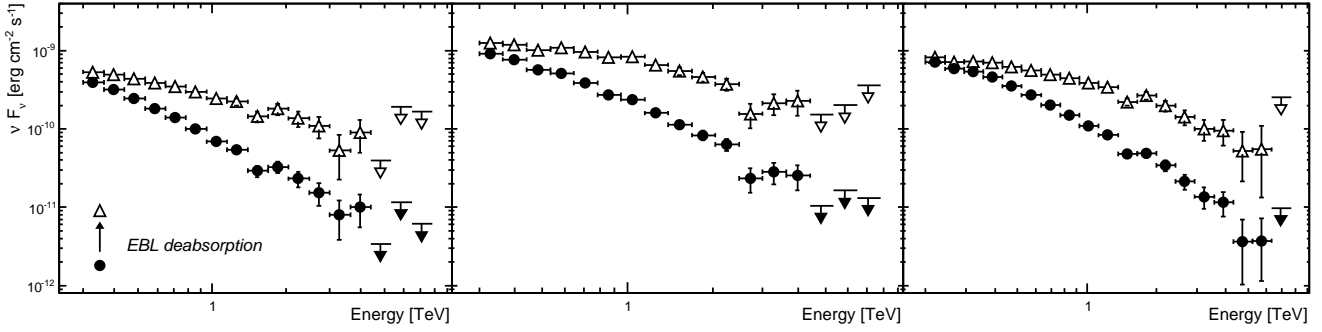}
     \caption{Selected VHE spectra: average spectra for the \texttt{T300-Low} (left)
      and \texttt{T300-High} 
     (middle) datasets (see Fig. \ref{fig:fluxindex300}), and 
     the average spectrum above 200 GeV (\texttt{T200}, right). 
     Open symbols corresponds to the spectra corrected for EBL absorption 
     as described in Sect. 4.1.2.}
     \label{fig:specs300}
\end{figure*}

%
\begin{table*}
\caption{Spectral fit of the measured VHE spectra extracted in different epochs (see Table 1). }
\label{tab:specs300}
\centering
\begin{tabular}{lccccccc}
\hline
\hline
\vspace*{-3mm}\\
PL fits  &   $\Phi_0$ & $\Gamma$   &       &          & $F_{0.3-3~\rm TeV}$ &     $\chi^2_r$ (d.o.f.) & $P_{{\chi}^2}$ \\
\vspace*{1mm}
         &  cm$^{-2}$s$^{-1}$TeV$^{-1}$&  &&          & erg cm$^{-2}$ s$^{-1}$ &         &      \\
\hline
\vspace*{-3mm} \\

\texttt{T400-Peak}   &2.20$\pm$0.06\e{-10} &3.46$\pm$0.04 &-&-& 1.35\e{-09} & 3.66 (11) & 3\e{-5}\\
\texttt{T300-High}   &1.36$\pm$0.03\e{-10} &3.36$\pm$0.03 &-&-& 7.88\e{-10} & 4.3 (12)  & 7\e{-7}\\
\texttt{T300-Low}    &4.89$\pm$0.15\e{-11} &3.51$\pm$0.03 &-&-& 3.09\e{-10} & 1.36 (12) & 0.18\\
\texttt{T200}        &7.46$\pm$0.12\e{-11} &3.25$\pm$0.01 &-&-& 4.06\e{-10} & 16 (16)   & 0\\
\texttt{T300-RXTE}   &4.78$\pm$0.30\e{-11} &3.53$\pm$0.07 &-&-& 3.07\e{-10} & 1.4 (11)  & 0.17\\

\hline
\vspace*{-3mm}\\
PL exp. cut. &$\Phi_0$       &   $\Gamma$ & $E_{\rm cut}$(TeV)  &       &  $F_{0.3-3~\rm TeV}$ &     $\chi^2_r$ (d.o.f.) & $P_{{\chi}^2}$ \\
\hline
\vspace*{-3mm} \\ 

\texttt{T400-Peak} &5.96$\pm$1.19\e{-10} &2.60$\pm$0.17 &1.19$\pm$0.25 &-& 1.28\e{-09} &0.38 (10) & 0.96\\
\texttt{T300-High} &2.72$\pm$0.36\e{-10} &2.86$\pm$0.09 &1.66$\pm$0.33 &-& 7.95\e{-10} &1.41 (11)  & 0.16\\
\texttt{T300-Low}  &7.08$\pm$1.25\e{-11} &3.26$\pm$0.12 &2.92$\pm$1.42 &-& 3.08\e{-10} &0.98 (11) & 0.47\\
\texttt{T200}      &2.12$\pm$0.23\e{-10} &2.65$\pm$0.06 &1.01$\pm$0.11 &-& 4.25\e{-10} &4.41 (15)  & 2\e{-8}\\
\hline
\vspace*{-3mm}\\
Log-parabolic &$\Phi_0$      &   $\Gamma$  &  $b$   &    &   $F_{0.3-3~\rm TeV}$ &     $\chi^2_r$ (d.o.f.) & $P_{{\chi}^2}$ \\
\hline
\vspace*{-3mm} \\ 

\texttt{T400-Peak}  &2.55$\pm$0.09\e{-10} &3.54$\pm$0.06 &1.05$\pm$0.20  &-& 1.25\e{-09} & 0.37 (10) & 0.96\\
\texttt{T300-High}  &1.46$\pm$0.04\e{-10} &3.53$\pm$0.05 &0.62$\pm$0.11  &-& 7.94\e{-10} & 1.35 (11) & 0.19\\
\texttt{T300-Low}   &4.98$\pm$0.16\e{-11} &3.66$\pm$0.07 &0.41$\pm$0.16  &-& 3.08\e{-10} & 0.76 (11) & 0.68\\
\texttt{T200}       &7.51$\pm$0.16\e{-11} &3.69$\pm$0.05 &0.78$\pm$0.07  &-& 4.29\e{-10} & 2.39 (15) & 0.002\\

\hline
\vspace*{-3mm}\\
Broken PL   & $\Phi_0$        &   $\Gamma_1$  &$\Gamma_2$   & $E_{\rm break}$(TeV)  &   $F_{0.3-3~\rm TeV}$ & $\chi^2_r$ (d.o.f.)  &$P_{{\chi}^2}$ \\
\hline
\vspace*{-3mm} \\ 

\texttt{T200}    &1.46$\pm$0.10\e{-10}  &2.73$\pm$0.05  & 3.60$\pm$0.04 & 0.42$\pm$0.02 & 4.31\e{-10} & 1.44 (14) & 0.12\\

\hline
\end{tabular}
\end{table*}


\begin{table*}
\caption{Spectral fit of the same VHE spectra given in Table \ref{tab:specs300},
  but corrected for EBL absorption (statistical errors only).}
\label{tab:deabs300}
\centering
\begin{tabular}{lccccc}
\hline
\hline
\vspace*{-3mm}\\
PL fits  &   $\Phi_0$ & $\Gamma$   &       &           $F_{0.3-3~\rm  TeV}$ &     $\chi^2_r$ (d.o.f.)  \\
\vspace*{1mm}
         &  cm$^{-2}$s$^{-1}$TeV$^{-1}$&  &&           erg cm$^{-2}$ s$^{-1}$ &              \\
\hline
\vspace*{-3mm} \\

\texttt{T400-Peak} &7.18$\pm$0.19\e{-10}&2.66$\pm$0.04 &-&3.01\e{-9}&5.13 (11) \\
\texttt{T300-High} &4.45$\pm$0.10\e{-10}&2.57$\pm$0.03 &-&1.81\e{-9}&4.95 (12) \\
\texttt{T300-Low}  &1.62$\pm$0.05\e{-10}&2.70$\pm$0.03 &-&6.88\e{-10}&1.13 (12) \\
\texttt{T200}	   &2.41$\pm$0.04\e{-10}&2.53$\pm$0.01 &-&9.74\e{-10}&7.88 (16) \\
\texttt{T300-RXTE} &1.59$\pm$0.10\e{-10}&2.72$\pm$0.07 &-&6.84\e{-10}&1.17 (11) \\

\hline
\vspace*{-3mm}\\
PL exp. cut. &$\Phi_0$       &   $\Gamma$ & $E_{\rm cut}$(TeV)&  $F_{0.3-3~\rm TeV}$ &     $\chi^2_r$ (d.o.f.) \\
\hline
\vspace*{-3mm} \\ 

\texttt{T400-Peak} &2.38$\pm$0.48\e{-9}&1.61$\pm$0.17 &1.00$\pm$0.18  &2.96\e{-9} &0.31 (10)\\
\texttt{T300-High} &9.14$\pm$1.15\e{-10}&2.04$\pm$0.09 &1.6$\pm$0.3   &1.83\e{-9} &1.50  (11)\\
\texttt{T300-Low}  &2.37$\pm$0.38\e{-10}&2.43$\pm$0.11 &2.9$\pm$1.4   &6.82\e{-10}&0.59 (11)\\
\texttt{T200}      &4.51$\pm$0.36\e{-10}&2.16$\pm$0.05 &1.74$\pm$0.24 &9.69\e{-10}&1.99 (15)\\

\hline
\vspace*{-3mm}\\
Log-parabolic &$\Phi_0$         &   $\Gamma$  &  $b$       &
$F_{0.3-3~\rm TeV}$ &     $\chi^2_r$ (d.o.f.)  \\
\hline
\vspace*{-3mm} \\ 

\texttt{T400-Peak}  &8.67$\pm$0.30\e{-10}&2.73$\pm$0.06 &1.23$\pm$0.20 &2.90\e{-9} &0.39 (10)\\
\texttt{T300-High}  &4.82$\pm$0.13\e{-10}&2.73$\pm$0.05 &0.61$\pm$0.11 &1.82\e{-9} &1.79 (11)\\
\texttt{T300-Low}   &1.66$\pm$0.05\e{-10}&2.83$\pm$0.07 &0.38$\pm$0.14 &6.81\e{-10}&0.51 (11)\\
\texttt{T200}       &2.44$\pm$0.01\e{-10}&2.79$\pm$0.04 &0.48$\pm$0.06 &9.67\e{-10}&1.76 (15)\\

\hline
\end{tabular}
\end{table*}

A study of the spectral variations was also performed
on the \texttt{T500} dataset, which allows the sampling 
of a wider time span and in particular of both the rise and decay phases
of the main $\gamma$-ray flare. However, the lower number statistics requires 
longer integration times,  yielding a lower time resolution.
The result is shown in Fig. \ref{fig:fluxindex500}, where spectra 
were extracted 
in 14 and 28 minutes bins.  The spectral variations follow the same pattern as for the 
\texttt{T300} spectra, both in time and in the flux-index relation.
No significant spectral changes are observed between the rising and decaying
part of the flare, with the possible exception of a hardening event that precedes
the peak of the $\gamma$-ray emission by $\approx$28 minutes. 

The study of the spectral shape in more detail requires higher event
statistics. To achieve this, the dataset was divided into similar spectral
states, namely a high and a low flux state 
(\texttt{T300-High} and \texttt{T300-Low}, respectively; see Fig. \ref{fig:fluxindex300}).
In addition, spectra were extracted in three other important epochs: 
a) around the peak of the $\gamma$-ray emission (see Fig. \ref{fig:fluxindex500}),
yielding a spectrum with a threshold of 400 GeV (\texttt{T400-Peak});
b) in the central five hours characterized by a threshold as low as 200 GeV (\texttt{T200} dataset);
c) in the epoch simultaneous with the \rxte exposure 
(\texttt{T300-RXTE} window, 44 minutes overall, see Table 1), 
where the combined X-ray spectrum can be measured over 2 decades in energy.

The results of the spectral fits are given in Table \ref{tab:specs300},
with a selection shown in Fig. \ref{fig:specs300}. 
The spectra present a significant curvature with respect to the pure power-law. 
The $\chi^{2}$ probabilities show that the latter is completely excluded 
in the high states, and is unlikely in the low state. 
The spectral curvature is generally well described either 
by a power-law model with an exponential cutoff around 1 TeV
(~$\Phi(E)=\Phi_{0}~E^{-\Gamma}~{\rm e}^{-E/E_{\rm cut}}$~), or a log-parabolic function
(~$\Phi(E)=\Phi_{0}~E^{-(\Gamma~+~b~{\rm log}(E))}$~). 
Most remarkably, the curvature of the spectrum is strongly variable with time.  
In particular, the curvature is more pronounced (i.e., the parameter $b$ is larger)
in the brightest state and decreases as the source dims. 
This represents direct proof that the curvature of the VHE spectrum in PKS\,2155--304
is also of intrinsic origin, inside the emitting region, 
and cannot be attributed entirely to $\gamma$-$\gamma$  absorption on the EBL  
or on any local external field  that is constant on the observed timescales.

Besides providing the widest energy coverage,
the \texttt{T200} spectrum
allows a direct comparison with the spectrum measured during the first exceptional 
flare on the night of July 27--28. 
The latter has the same energy threshold (200 GeV) and is well described by a broken
power-law (~$\Phi(E)=\Phi_{0}~E^{-\Gamma_1}$ for $E<E_{\rm break}$ and 
$\Phi(E)=\Phi_{0}~E_{\rm break}^{\Gamma_2-\Gamma_1}E^{-\Gamma_2}$ for
$E>E_{\rm break}$~)
with $\Gamma_1=2.71\pm0.06$, $\Gamma_2=3.53\pm0.05$, and $E_{\rm break}=430\pm22$ GeV. 
Fitting this function to the \texttt{T200} spectrum 
yields almost identical results, of the same break energy, slopes
and change in spectral index by $\Delta\Gamma\simeq0.9$ (Table
\ref{tab:specs300}). This shows that the source was in a similar state, 
even though the overall average normalization is about $\sim$30\%
lower than two nights before. As for the July 27--28 night, the \texttt{T200} spectrum
is not well fitted by a power-law model with exponential cutoff or a
log-parabolic function (F-test $>$99\% compared to the broken power-law). 
Both functions underestimate significantly the $\gamma$-ray flux at higher energies.

\subsubsection{Correction for intergalactic $\gamma$-$\gamma$ absorption}
The VHE $\gamma$-ray emission from extragalactic sources is expected to be attenuated 
by photon-photon interactions with the EBL
photons in the optical-to-IR waveband. 
The energy dependence of the optical depth 
-- which is determined by the spectrum of the diffuse background -- 
causes a general steepening of the emitted $\gamma$-ray spectrum,
more or less severe according to the energy band considered  
\cite[see e.g.,][and references  therein]{felixicrc}.
At redshift $z$=0.116, this effect is substantial
and must be taken into account to study the true energy output, spectral properties, 
and location of the Compton peak in the SED \citep{2155mwl,nature_ebl}.
Source diagnostic based on flux or spectral variability, instead, is unaffected, 
since intergalactic absorption is a constant factor for all purposes
related to blazar variability (the diffuse background varies only on cosmological timescales).

The EBL waveband that affects the observed VHE band the most
is dominated by the direct starlight emission.
To correct for EBL absorption, as reference we adopted  the 
model of the EBL spectral energy distribution 
by \citet{franceschini}, which is based on the emission from galaxies. 
This  model takes into account the most 
recent results on galaxy properties and evolution and is consistent
with both the lower limits from source counts  -- in the UV-optical 
\citep{madaupozzetti} as well as near--mid infrared waveband \citep{fazio,dole06} --
and with the upper limits derived from the TeV spectra of high-redshift blazars
\citep{nature_ebl,hess0347,hess0229,hegra1426}. 
It is similar in shape to both the model by \citet{primack} and the
``low--IR'' calculation by \citet{kneiske2004}.
The spectra were corrected by applying the optical depth 
calculated for the average observed photon energy in each energy bin.

However, it is important to recall that 
a significant uncertainty in the SED of the EBL 
still remains, since it could be both lower and higher than assumed:
either down to the absolute lower limits given by  HST galaxy counts \citep{madaupozzetti} 
\citep[as in the model by][]{primack}, or up to the upper limits given by TeV blazars
\citep{nature_ebl}. To estimate this uncertainty in both shape and normalization, 
we also used the shape of the model by \citet{primack}, rescaled to these two levels.
In the energy range around the starlight peak (1--3 \micron),
the residual uncertainty in the EBL absolute normalization is of the order of 50\%  
(from $\sim$8 to $\sim$12 \nw at 2.2 micron, while our reference model gives 9.4 \nw).
This translates into a systematic uncertainty of the order of $\Delta\Gamma\simeq \pm0.2$ 
in the reconstructed $\gamma$-ray spectrum.  
Namely, the reconstructed spectra (which we call ``intrinsic'') discussed in the following
Sects. can actually be up to $\sim$0.2 steeper or harder than indicated.
When relevant, we take this systematic uncertainty into consideration, but
as we show in the following, it does not change the main properties of the  $\gamma$-ray spectrum 
and Compton peak frequency  of PKS\,2155--304.

\begin{figure*}[t]
\centering
   \includegraphics[width=9cm]{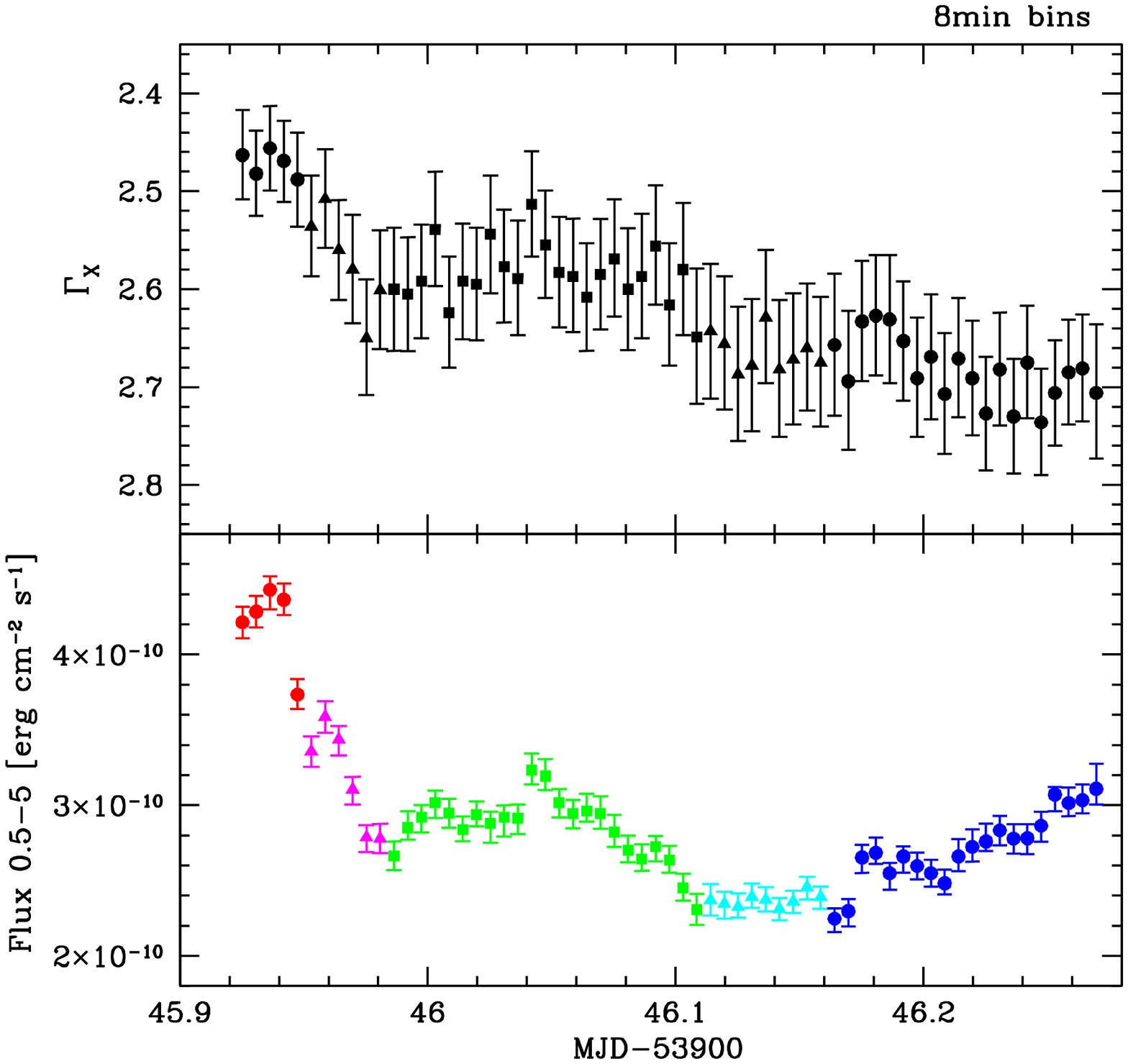}\hspace{1cm}
   \includegraphics[width=8cm]{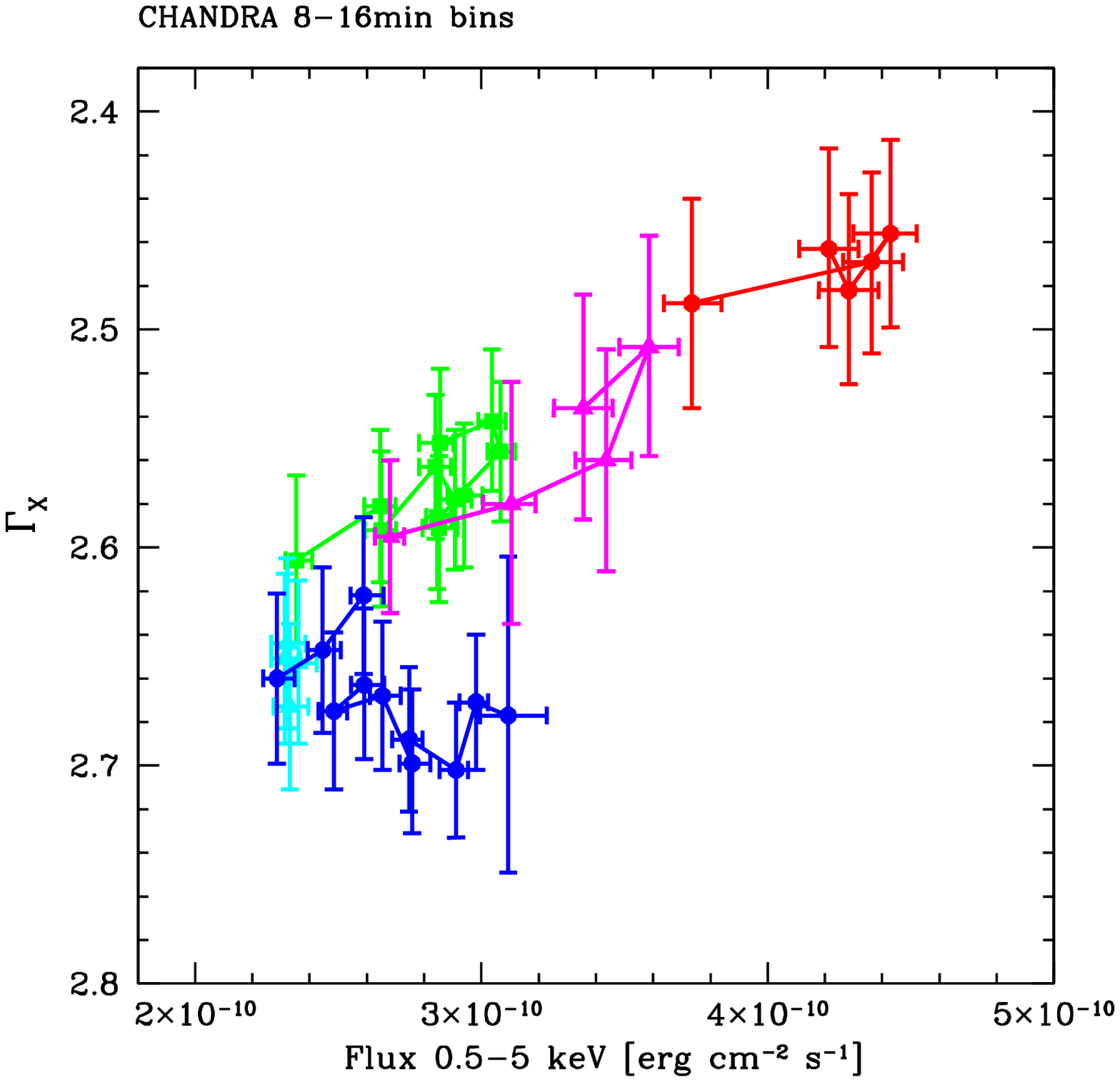}
      \caption{Evolution of the X-ray spectrum with time, for the whole
      {\chandra} exposure. Left panel: each 8-min bin is fitted with
     a single power-law  model plus galactic absorption. 
     The \hess window ends at \mjd=46.16. 
     Left lower panel: the colors and symbols (in time sequence: red circles, magenta triangles, green squares, 
     cyan triangles, blue circles)
     mark the different zones for the index-flux correlation shown in the
     right panel. Right panel: for clearer visibility, the last two intervals 
     (cyan triangles and blue circles) are further binned into 16-minute spectra.
     There is a clear ``harder when brighter'' trend in the decaying
     phase of the first branch (corresponding to the VHE window). 
     The behaviour changes in the successive rising phase:
     the X-ray spectrum continues to soften while the flux increases,
     drawing a counter-clockwise pattern.
     } 
     \label{xloops}
\end{figure*}

\subsubsection{Absorption-corrected $\gamma$-ray spectra}
For the power-law spectra measured on short timescales 
(as given in Fig. \ref{fig:fluxindex300}),
the intrinsic spectra are again well fitted by a power-law model
with a slope that is typically harder by $-$0.8
(namely, $\Gamma_{\rm int}=\Gamma_{\rm obs}-0.8$).
The results of the fits to the spectra with higher event statistics  
are  provided in  Table \ref{tab:deabs300}.
Even after correction for the steepening induced by EBL absorption,
the $\gamma$-ray spectra show clear evidence of curvature,
and the power-law model is excluded with high confidence for all states 
but the \rxte epoch (which has the lowest exposure).
The spectral curvature is well described by either the log-parabolic function or 
a power-law model with an exponential cutoff around 1-2 TeV,
for all spectra, including the \texttt{T200} dataset.

When a spectrum shows a relatively uniform curvature 
as in this case, however,  the log-parabolic model is generally preferable.
It has the advantage of providing a more direct measure of the curvature 
in the true observed band, whereas the exponential cutoff model tends 
to match a given curvature in the observed passband 
by using a specific section  of its cutoff region, and pushing the power-law
component outside the actual observed range.
This often yields artificial values for the slope, which are typically too hard.
The log-parabolic fit allows also a straightforward estimate of the location of the 
SED peak  (E$_{peak}$, defined by $\Gamma({\rm E}_{peak})=2$)
from the curvature itself, with a minimum of free parameters.
To this aim we used the functional form described in \citet{tramacere},
where $b$ and E$_{peak}$ are the independent free parameters instead of $b$ and $\Gamma_{\rm 1 TeV}$.
The comparison of the curvatures among different states and between
the synchrotron and IC components can also provide important clues about the source emission 
regime (Thomson or KN) and the acceleration mechanisms \citep{massaro3}.

By comparing  the spectra in the 3 different flux states
(\texttt{T400-Peak}, \texttt{T300-High} and \texttt{T300-Low}),
one can see that the curvature changes significantly along the night
(at a confidence level $>99.99$\%) and so does the Compton peak energy.  
A clear trend emerges: both the curvature parameter $b$ and E$_{\rm peak}$ 
increase with the VHE flux.
At the maximum of the VHE flare, the spectrum is strongly curved 
($b=1.2\pm0.2$), with the Compton peak estimate at E$_{peak}=500\pm50$ GeV. 
As the flux decreases,  the curvature flattens ($b=$0.62 to 0.35),
while the IC peak shifts to lower energies (E$_{peak}=260\pm35$ GeV to
E$_{peak}=70\pm50$ GeV, respectively).
A lower/higher EBL level does not change these results substantially:
a higher level yields similar curvatures ($b$=1.4, 0.76, and 0.48, respectively)
and slightly higher IC peak energies because of the generally harder spectra
(E$_{peak}$= 580, 370, and 180 GeV, respectively). 
It is important to recall that the absolute value of the curvature $b$ 
depends on the particular choice of the EBL spectrum used, but not the trend itself.
This trend is opposite to what is generally observed and expected
for the synchrotron emission in TeV blazars. For example,
in Mkn\,421 the curvature $b$ decreases as both E$_{peak}$ and the flux increase
\citep{massaro1}.
We also note that the spectral index $\Gamma$ 
does not correlate (and possibly anti-correlate) with the curvature $b$,
in contrast to what is observed in the X-ray band for this (see next Sect.)
and other HBL \citep{massaro1}.

\subsection{X-ray spectra}
\label{sect:xspectra}
The {\chandra} spectra were extracted 
both by a uniform sampling of the whole exposure,
in different time bins (2-4-8-16 minutes),  and strictly coincident with the VHE time bins.
The spectra were  all fitted using an equivalent hydrogen column density
fixed at Galactic values ($1.69\times 10^{20}$cm$^{-2}$) 
with different  source models. 
On short integration times ($<1$ hr), a single power-law model 
provides statistically good fits for all datasets,  
while evidence of curvature is found only when larger exposures are considered 
or a wider energy band is available (for example including the \rxte data).

\begin{figure*}[th]
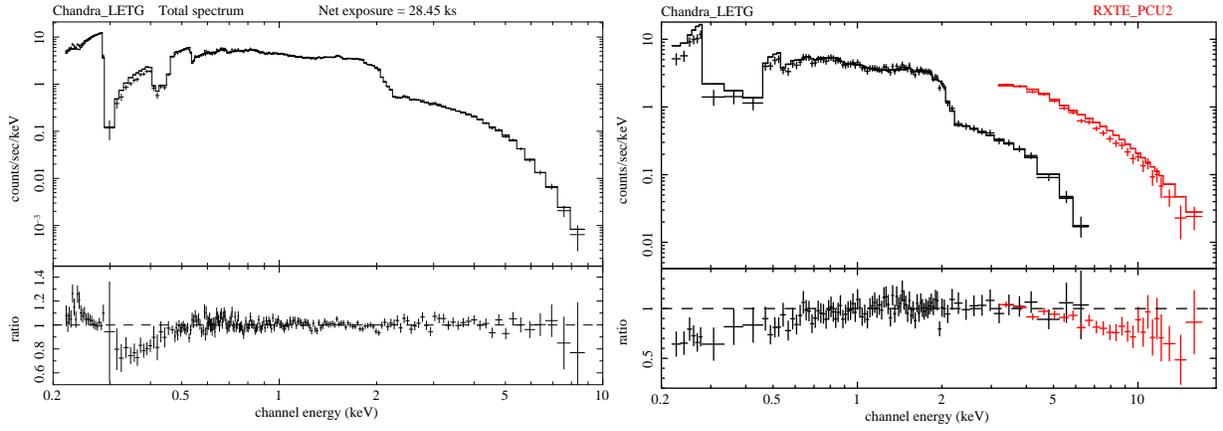

\centering
 \includegraphics[angle=-90,width=8cm]{figures/letg_total_bkpow2.ps}
 \includegraphics[angle=-90,width=8cm]{figures/cxte_fitpow2.ps}
 \caption{Left panel: {\chandra} spectrum for the total exposure, fitted with a broken power-law
 plus galactic column density. The additional absorption feature in the 0.3-0.4 keV range is due 
 to the contaminants on the ACIS optical blocking filter, not yet fully accounted by the calibration. 
 It can be accounted for with a simple edge model at 0.31 keV and $\tau_{\rm max}$=0.4 (see text).
 Right panel: the simultaneous \chandra+\rxte spectrum in the common 44-min window. 
 The plot corresponds to a single power-law with galactic absorption,
 and the data/model ratio shows the clear evidence of curvature (fit parameters are given in 
 Table \ref{xtabula}).
 The \rxte/{\chandra} normalization is fixed at 1.08, as derived from the fit in the
 overlapping energy range (3--7 keV).}
      \label{xspectra} 
\end{figure*}

\begin{table*}[th]
\caption{Fit of the X-ray spectra simultaneous to the VHE data. }
\centering
\begin{tabular}{lcccccccc}
\hline
\hline
Broken-PL fits  & Exposure & Band & $\Gamma_1$ & $E_{\rm break}$    & $\Gamma_2$ 
&$F_{0.5-5~\rm keV}$  & $F_{2-10~\rm keV}$ & $\chi^2_r$ (d.o.f.) \\
\vspace*{1mm}
                 & $ks$ &    keV     &           & keV             &          &  
\multicolumn{2}{c}{erg cm$^{-2}$s$^{-1}$} &  \\
\vspace*{-3mm}\\
\hline
\vspace*{-3mm} \\
LETG \texttt{T300-High}& 3.9  & 0.2--8 & $2.35\pm0.03$ & $1.00\pm0.07$ &  $ 2.60\pm0.02$ & 3.88\e{-10} & 1.45\e{-10} & 0.65 (204)
\vspace*{1mm}\\  
LETG \texttt{T300-Low}& 9.4  & 0.2--8 & $2.41\pm0.02$ & $0.95\pm0.06$ &  $ 2.71\pm0.02$ & 2.78\e{-10} & 9.19\e{-11} & 0.74 (204)
\vspace*{1mm}\\  
LETG  \texttt{T200}   & 14.1  & 0.2--9 & $2.39\pm0.01$ & $0.95\pm0.04$ &  $ 2.68\pm0.01$ & 3.05\e{-10} & 1.05\e{-10} & 0.87 (204)
\vspace*{1mm}\\  
LETG  \texttt{T300-RXTE}& 2.6  & 0.2--20 & $2.56\pm0.02$ & $2.72\pm0.22$ &  $ 2.98\pm0.04$ & 2.79\e{-10} & 9.11\e{-11} & 0.69 (117) 
\vspace*{1mm} \\
\hline
\vspace*{-1mm} \\
Log-parabolic fits &Exposure &Band & $\Gamma$    &    $b$  &             &$F_{0.5-5~\rm keV}$
 & $F_{2-10~\rm keV}$ & $\chi^2_r$ (d.o.f.) 
\vspace*{1mm} \\
\hline
\vspace*{-3mm} \\
LETG \texttt{T300-High}& 3.9   & 0.2--8  & $2.48\pm0.01$ & $0.18\pm0.03$ &  -  &  3.88\e{-10} & 1.41\e{-10} & 0.69 (205) 
\vspace*{1mm}\\  
LETG \texttt{T300-Low}& 9.4   & 0.2--8  & $2.57\pm0.01$ & $0.21\pm0.02$ &  -  &  2.77\e{-10} & 8.89\e{-11} & 0.83 (205) 
\vspace*{1mm}\\  
LETG  \texttt{T200} & 14.1   & 0.2--9  & $2.55\pm0.01$ & $0.21\pm0.01$ &  -  &  3.05\e{-10} & 1.01\e{-10} & 1.02 (205) 
\vspace*{1mm}\\ 
LETG  \texttt{T300-RXTE}& 2.6  & 0.2--20 & $2.56\pm0.01$ & $0.25\pm0.02$ &  -  &  2.81\e{-10} & 8.92\e{-11} & 0.60 (118) 
\vspace*{1mm} \\
\hline
\end{tabular}
\label{xtabula}
\end{table*}    

The time evolution in the X-ray spectrum during the entire {\chandra} 
pointing --which extends few hours beyond the end of the \hess observation--
is shown in Fig. \ref{xloops}.
There is a clear trend of ``harder-when-brighter'' behaviour 
in the first part of the dataset, corresponding 
to the decaying phase of the main VHE flare. 
This behaviour is also followed by the small-amplitude flares, 
whose paths in the flux--index plane  overlap
tightly with the overall trend of the decaying phase (Fig. \ref{xloops} right panel).

However, the relation changes in the last part of the observation (\mjd $>$46.16):
as the X-ray flux starts to increase again,  the spectral index continues to soften. 
This ``softer-when-brighter'' behaviour in the rising phase of a new flare 
reveals a change in conditions for the emitting region. 
It is indicative of a slow acceleration/injection process, 
whose timescale is comparable with the other timescales of the system  ($t_{acc}\approx t_{cool}$).
The information about the flare then propagates from lower to higher energies as particles 
are gradually accelerated \citep{kirk98,ravasio421}.
If the optical variations are indeed associated with the flaring zone,
the optical data would support this scenario as well, exhibiting
increasing flux just before the X-ray rise
at the end of the {\chandra} observation.
Together, the two patterns of the X-ray data  draw part of a counter-clockwise 
loop in the flux--index plane (Fig. \ref{xloops}).

Fig. \ref{xspectra} shows the spectra of both the total {\chandra} exposure and 
the \rxte simultaneous window. 
The results of the fits performed on the \hess-simultaneous datasets 
are  given in Table \ref{xtabula}.
All datasets correspond to strictly simultaneous windows except 
for the \texttt{T300-High} spectra, for which the X-ray window
does not include the first $\sim$10 minutes of the 1.3-hrs VHE window above 300 GeV.
Since there are no significant spectral changes at VHE in that window,
the VHE spectrum can be considered to accurately represent the 
$\gamma$-ray spectral shape in the X-ray window.

For all spectra in Table \ref{xtabula}, there is clear evidence of curvature, 
and the single power-law model is rejected with high confidence (F-test$>99.99\%$). 
The spectra show a continuous steepening towards higher energies up to $\sim$20 keV,
which is well represented by both a broken power-law and log-parabolic models. 
A power-law with an exponential cutoff is excluded as well ($P_{{\chi}^2}<$0.009)
for the spectra with the highest statistics 
(\texttt{T200} and total spectrum).
The drop rate in the cutoff region is significantly slower than 
${\rm e}^{-E/E_{cut}}$
and  also slightly slower than 
${\rm e}^{-(E/E_{cut})^{1/2}}$, 
as indeed expected for the synchrotron emission of a particle distribution
with an exponential cutoff \citep{felix2000}.

As is clear from Fig. \ref{xloops} and Table \ref{xtabula},
during all times the X-ray spectrum of PKS 2155--304 remains steep, 
with a convex shape and no signs of flattening at high energies
\citep[as instead found in {\xmm} observations performed in November 2006;][]{foschini_rem,zhang08}.
This means that the peak of the synchrotron emission has not entered 
the observed energy range at any time, and that there is no sign of the possible emergence
of the IC component in the hard X-ray band.

It is interesting to compare the curvature parameters and SED peak 
location given by the log-parabolic fits.  
Both the spectral index and the curvature increases (slightly)  as the flux decreases. 
This is also corroborated by the fits of spectra extracted in even shorter intervals at 
the two extreme of the X-ray flux range (namely \texttt{T300-Xmax} and \texttt{T400-Xmin}).
The log-parabolic fit yields $\Gamma=2.45\pm0.02$ and $b=0.16\pm0.04$
versus $\Gamma=2.64\pm0.01$ and $b=0.24\pm0.04$, respectively, for an integrated
flux  $F_{0.5-5 {\rm keV}}=$4.33 and 1.65\ergs{-10}. 
The change in the two spectral parameters, however, is such that the estimate 
of the location of the SED peak remains basically constant:
for all spectra, the synchrotron $E_{peak}$ falls within the range 40--50 eV 
(with a typical 1-$\sigma$ statistical error of $\pm20$ eV).
In contrast to the behaviour in $\gamma$-rays, in the X-ray band 
the photon index shows a positive correlation with the curvature $b$,
as typically observed for example in Mkn 421 \citep{massaro1}.
The absolute values of the curvature are similar to those found 
for most other HBL \citep{massaro2008}.

\begin{figure}[t]
\centering
   \includegraphics[width=9cm]{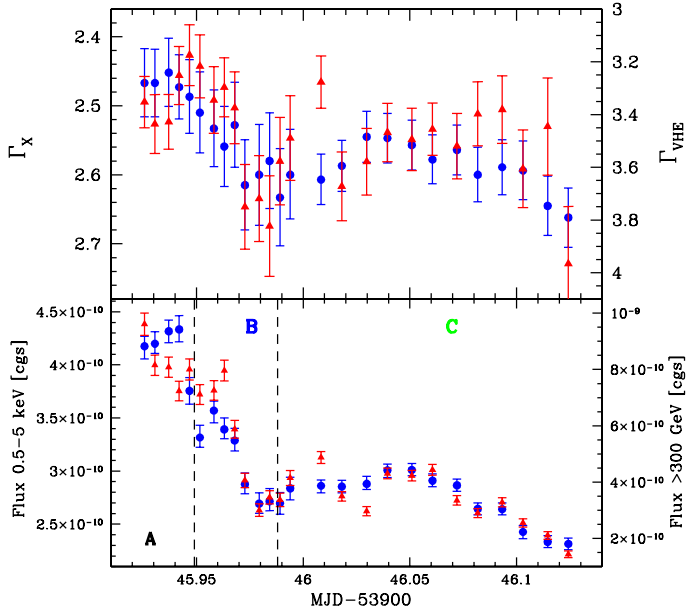}
     \caption{Plot of the simultaneous  spectral and flux variability
in the X-ray (blue circles) and VHE (red triangles) bands, in the \texttt{T300-X} window.
The spectra have been extracted in the same time-bins,
with integration times of 7 and 14 minutes (before and after \mjd 46.0, respectively).
The vertical scales are different (left: X-ray values; right: VHE values).
Upper panel: the scale of the VHE photon indices is 2.6 times the X-ray scale.
The VHE indices refer to the observed spectra; the corresponding values 
after correction for EBL absorption can be obtained as $\Gamma_{intr}=\Gamma_{\rm VHE}-0.8$.
Lower panel: integrated energy fluxes in the same units of erg cm$^{-2}$ s$^{-1}$.
The range of the VHE scale is the \emph{cube} of the X-ray range 
(15.3$\times$ versus 2.5$\times$).
The vertical lines mark the three time-zones (A, B, and C) referred to in the text 
and in Fig. \ref{fxfg}.
}
     \label{714min}
\end{figure}

\begin{figure}[t]
\centering
   \includegraphics[width=9cm]{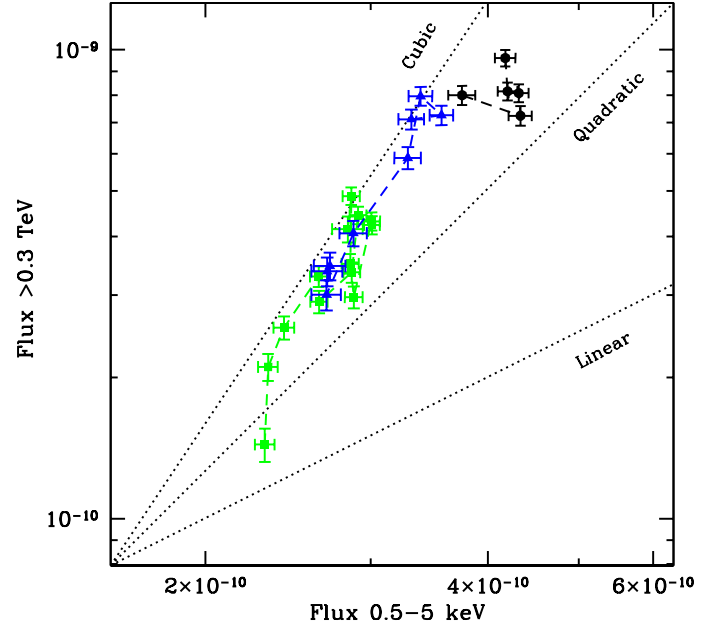}
     \caption{Plot of the $\gamma$-ray vs. X-ray flux correlation, in a log-log diagram
     and units of erg cm$^{-2}$ s$^{-1}$.
     Different markers correspond to the three different intervals shown in Fig. \ref{714min}:
     A (black circles), B (blue triangles)  and C (green square).
     For visual reference, the dotted lines show
     three different slopes of the relation $F_{\gamma}\propto F_x^{\beta}$.
     For clearer visibility, the Y-scale corresponds to the \emph{square}
     of the X-scale, so that a quadratic relation has the slope of 1. 
     The best-fit values are given in Table \ref{fxfgtable}.
     }
     \label{fxfg}
\end{figure}

\section{X-ray vs. TeV correlations}
\label{sect:xtev}
\subsection{Spectral variability}
The exceptional $\gamma$-ray brightness observed in this night, 
coupled with the sensitivity of the \hess array and the continuous coverage provided 
by {\chandra}, 
allows the emission in the two bands to be compared
with unprecedented time resolution in the spectral domain as well.

Fig. \ref{714min} shows the simultaneous flux and spectral properties 
measured in 7 and 14 minute time bins, in the \texttt{T300-X} window.
Inside this window, the VHE spectral index can be well constrained ($\pm0.1$) over 
approximately a decade in energy (0.3--2 TeV).
The different binning was chosen to achieve comparable S/N ratio during the night, 
and as a good compromise between spectral determination at VHE and time resolution
(see Sect. 4.1).
Both X-ray and VHE spectra have been extracted in exactly the same time bins.
The spectra were fitted with a single power-law model, which provides 
a good fit for each time bin. The integrated energy fluxes  were 
calculated using the specific spectral value measured in each bin.  

Two general properties can immediately be  noted.
The first is that the VHE emission shows a definite correlation with the X-ray emission 
not only in flux but also spectrally. 
The spectral evolution  follows the same overall pattern in the two bands, although 
with different amplitudes.  The correlation coefficient between the X-ray and $\gamma$-ray 
spectra is $r=0.65$,  with a probability $P<0.1$\% of a chance correlation. 

The second property, as previously illustrated by Fig. \ref{nufnu},
is that the source shows amplitude variations much larger in $\gamma$-rays than in X-rays.
This is now evident also for the spectra, although not as dramatically as for the flux:
the spectral variation at VHE 
is about 3 times the variations in the X-ray band  
($\Delta\Gamma_{\rm VHE}\approx0.65$ vs. $\Delta\Gamma_{\rm X}\approx0.21$).

The combination of such large-amplitude variability at VHE with
correlated but small-amplitude variations in X-rays yields one of the 
most striking features of this dataset: 
the VHE flux \emph{varies  more than quadratically}
with respect to the X-ray flux \emph{during a decaying phase}.

\subsection{Cubic relation between X-ray and TeV fluxes}
Fig. \ref{fxfg} shows the VHE flux as a function of the X-ray flux, 
in a log-log diagram. 
The data are divided into three subsets (``A'', ``B'', and ``C'') 
corresponding to three characteristic epochs: 
epoch ``A'' covers the first 35 minutes of the
simultaneous window; epoch ``B'' corresponds to the full rise and decay phases 
of the small-amplitude flare at \mjd 45.96; epoch ``C'' covers the remainder of the dataset, 
corresponding to the broader flare (see Fig. \ref{714min}).

The data were fitted with a linear relationship in the log-log space 
($F_{\gamma}\propto F_{\rm X}^{\beta}$), for the total dataset and in each subset separately.
The results are reported in Table \ref{fxfgtable}, together with the fits of the same 
datasets with a finer sampling (4-minute bins).

The $\gamma$-ray flux traces the variations in the X-ray flux
far more than quadratically, namely approximately as $F_{\rm VHE}\propto F_{\rm X}^{3}$.
With the exception of the first few minutes of epoch A,
all points lie on a narrow path, also during 
the small flare on \mjd 45.96 (epoch B).
In epoch A, instead, the two emissions do not seem to correlate. 
The lack of correlation is limited to a few points and in a short interval;
however, one may speculate whether this is caused by the rapid, low-significance 
structures in the VHE light curve, 
which are not evident in X-rays (as discussed in Sect. 3.3).

One of these subflares is indeed present in epoch A 
around MJD$_0 =45.925$ (see Fig. \ref{fig:xg300}), and its exclusion 
does bring the $\gamma$-ray emission  more in line with the X-ray pattern.
This type of $\gamma$-ray subflares on top of correlated emissions
have been recently envisaged by \citet{ggmagnetic}, but more sensitive 
instruments are required to draw any conclusions.

In the epochs B and C,  the cubic correlation is obtained by considering both 
the two zones separately and together,  for which an even steeper slope of $\beta=3.35$ is obtained.
This cubic correlation is robust with respect to the inclusion 
or exclusion of single data points and, in particular,
does not depend on the lowest VHE point.

The flux-flux correlation is plotted  using the same units for both X-ray and VHE bands,
namely the integrated energy fluxes.
Compared to previous studies, which used  the observed 
event rates, our approach is more consistent and it is allowed by the good spectral 
determination in each time bin. 
However, we remark that the measured relation does not depend 
significantly  on the particular approach used \citep[see discussion in][]{fossati08}:
a cubic relation is also obtained  by using  the event rates, 
both in the VHE band (as photons cm$^{-2}$ s$^{-1}$) and in the X-ray band (count rate).

The correlation found in the total dataset is less steep than in each subset,
while in the B+C epoch it is slightly steeper than for each of B and C separately.
This can be caused by a possible shift in 
the flux-flux paths among different datasets/epochs.
While individual paths still obey a specific steep trend, 
taken together they can produce a flatter (or steeper) envelope. 
This effect was indeed observed in Mkn 421 \citep{fossati08},
when considering data for different days.

During the decaying phase of a flare, 
a cubic relation between the $\gamma$-ray and X-ray fluxes 
is not easy to explain even for a source in Thomson condition, 
if both fluxes sample the emission beyond the respective SED peaks
\citep{katar05}.
It is the first time that such a steep slope has been observed in the history of 
the X-ray/TeV correlation studies,
though indication of a super-quadratic relation was recently
reported  for Mkn\,421 \citep{fossati08}, during single flares.

\begin{table}[t]
\caption{Values of the slope $\beta$ of the correlation $F_{\gamma}\propto F_x^{\beta}$.
Fluxes integrated over each respective energy band (0.5-5 keV and 0.3-3 TeV)
and in strictly simultaneous bins. The parameter is shown for fits performed with two different binnings
(4 min and 7-14 min) and over three different intervals, as shown in Fig. \ref{714min}. }
\centering
\begin{tabular}{lcc}
\hline
\hline
Datasets     &    4-min bins &	7-14min bins     \\
\hline
all       &   $2.21\pm0.05$  &  $2.25\pm0.05$      \\
\hline
A         &     no corr.     &     no corr         \\
B         &   $2.72\pm0.17$  &  $3.18\pm0.18$       \\
C         &   $2.83\pm0.17$  &  $3.14\pm0.18$     \\
B+C       &   $3.13\pm0.11$  &  $3.35\pm0.11$      \\ 
\hline
\end{tabular}
\label{fxfgtable}
\end{table}    

\begin{table}[t]
\caption{Values of the slope $\beta$ of the correlation $F_{\gamma}\propto F_x^{\beta}$,
for fluxes integrated over different energy bands, as indicated.
The 3-15 keV flux is obtained by extrapolation of the  power-law model in the \chandra~ passband.  
Fit of the dataset ``B+C'' in the 7-14 minutes binning. }
\centering
\begin{tabular}{lcc}
\hline
\hline
              &  \multicolumn{2}{c}{VHE bands}       \\
\hline
X-ray bands   &  0.3-0.7 TeV     &  $>$0.7 TeV    \\
\hline
\vspace*{-2mm} \\
0.5-5 keV     &   $2.91\pm0.12$  &  $4.11\pm0.27$       \\
3-15 keV      &   $1.87\pm0.08$  &  $2.70\pm0.20$    \\
\hline
\end{tabular}
\label{fxfgbands}
\end{table}

These correlations have been studied so far
mainly with the \rxte-PCA instrument, which samples higher energies 
than those observed here. If the spectrum changes with the flux,
and  with a ``harder-when-brighter'' behaviour as in this case,
the amplitude of the variations changes with energy, 
and thus the slope of the correlation can depend on the observed band.

To quantify this effect, we also investigated the flux-flux relation
by extrapolating the {\chandra} spectrum in the \rxte band 
(namely, integrating the best-fit model in the 3--15 keV range), 
and dividing the VHE range into soft and hard bands 
(0.3--0.7 and $>$0.7 TeV, respectively).
The result is shown in Table \ref{fxfgbands}.
The same cubic correlation observed in {\chandra} and \hess translates into a 
quadratic relation (in fact quite similar to that observed in Mkn 421) 
between the hard X-ray and soft VHE bands, while
an even steeper slope ($\beta>4$) 
is obtained between the soft X-ray band and hard VHE band.
We note however that the cubic relation is not simply the effect of 
a pivoting type of variability seen in  different energy bands.
The VHE band is closer to the IC peak (i.e the pivoting point)
than the X-ray band to the synchrotron one, thus the X-ray variations 
should be larger than the VHE ones, 
in contrast to what is observed. 
It is the entire IC peak that has actually varied far
more than the synchrotron peak, as shown also by the nearly constant and cubic  
values of the correlation  between corresponding bands 
(i.e., soft-soft and hard-hard, see Table 3).

Comparing these 2006 data (low state) with the 
X-ray/TeV campaign of 2003 \citep{2155mwl},
the overall brightening 
in the X-ray (2--10 keV) and VHE  
($>$300 GeV) bands is similar (a factor of 2.7 and 5, respectively), 
and corresponds to a relation 
$F_{\gamma}\propto F_{\rm X}^{1.6}$ between the two epochs.
Thus PKS\,2155--304 has varied its overall synchrotron and IC luminosity 
sub-quadratically on very long timescales, but super-quadratically
on intra-night timescales, at least during this major flaring event.

\begin{figure*}[t]
\centering
   \includegraphics[width=9cm]{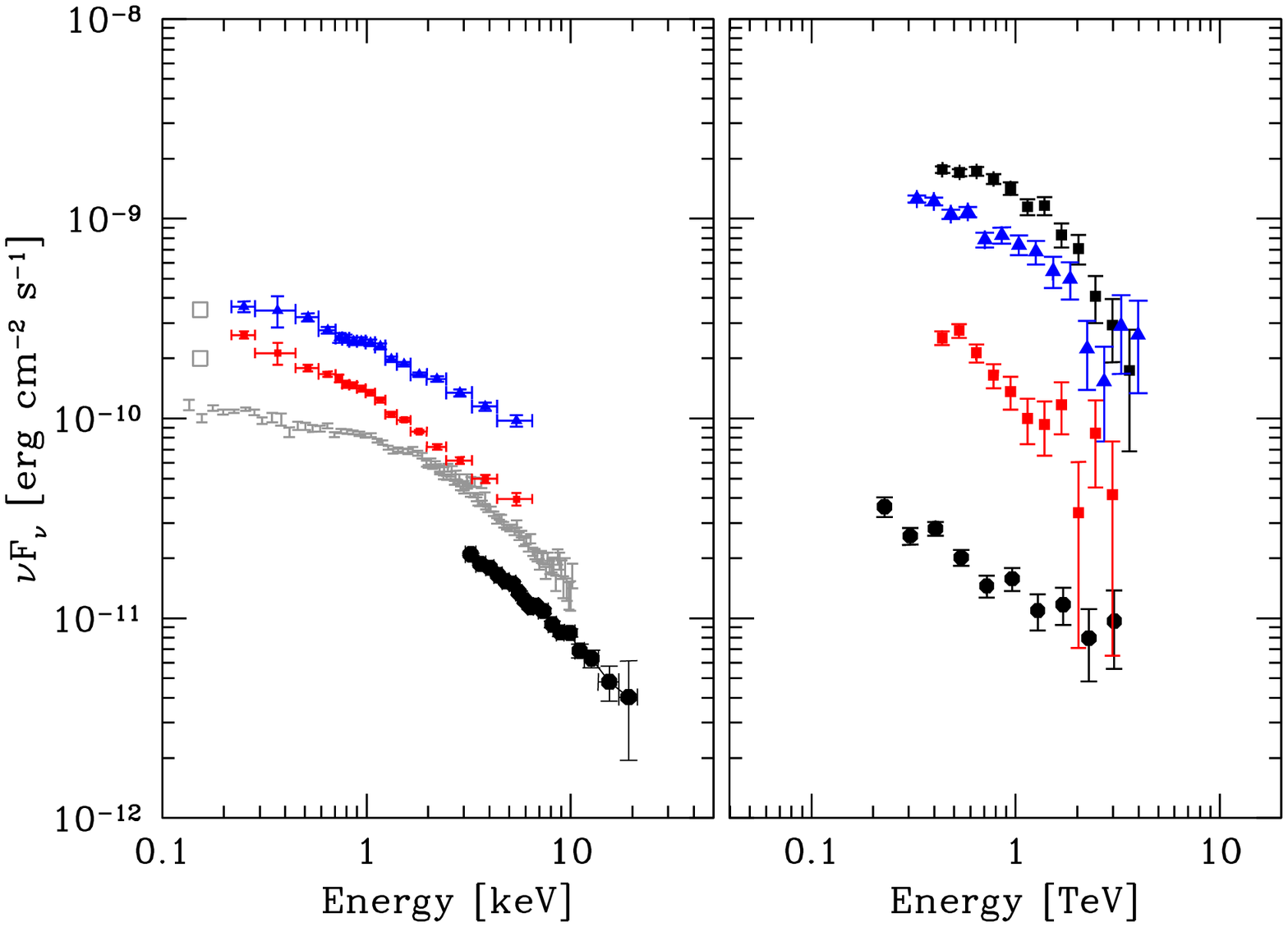}
   \includegraphics[width=9cm]{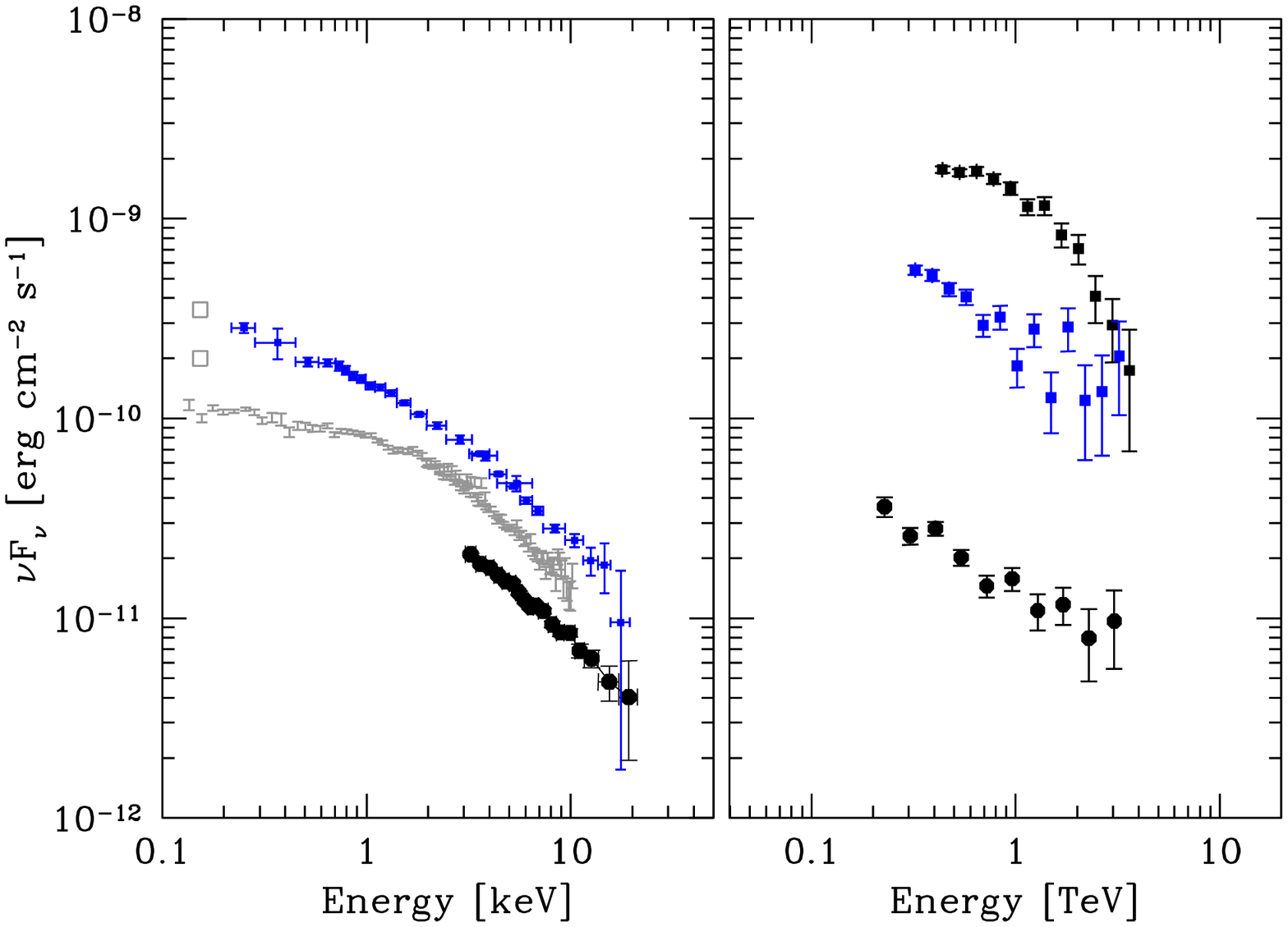}
   \caption{Gallery of simultaneous X-ray and VHE SED-pairs 
  in different states along the night. Left panel: 
  maximum (\texttt{T300-Xmax} blue triangles) and minimum (\texttt{T400-Xmin}, red squares) 
  states (see Table 1 and Fig. 2).   Right panel: simultaneous 
   \chandra+\rxte and \hess spectra (\texttt{T300-RXTE}, blue squares).
   For comparison, the following data are also plotted: 
   the VHE spectrum at the peak of the $\gamma$-ray flare (\texttt{T400-Peak}; upper black squares); 
   the multiwavelength \rxte-\hess campaign in 2003 \citep[lower black circles;][]{2155mwl};
   historical EUVE flux and BeppoSAX spectrum \citep[from][grey points]{chiappetti99}.
   All VHE data are corrected for EBL absorption according to \citet{franceschini} 
   (see Sect. 4.1.2).  
   }
   \label{sedpairs}
\end{figure*}

\section{Spectral Energy Distributions}
%
\subsection{X-ray/$\gamma$-ray spectra pairs}
To highlight the evolution of the SED during the night,
a gallery of selected pairings of simultaneous X-ray and VHE spectra 
is shown in Fig. \ref{sedpairs}. 
The scales on both axes are kept the same for both X-ray and $\gamma$-ray energies,
to enable visually a correct comparison of the spectral slopes.
For reference, a selection of historical observations is also plotted, 
in particular the data from the first X-ray/TeV multiwavelength campaign 
performed on PKS\,2155-203 in October 2003 \citep{2155mwl}. 
These data (obtained with \rxte and HESS) correspond to one of the historically lowest 
states ever observed from this object.

The panels show the SED-snapshots taken in 
the brightest/hardest state (\texttt{T300-Xmax}), in the faintest/softest state (\texttt{T400-Xmin}),
and simultaneously with the \rxte pointing (\texttt{T300-RXTE}).
In addition, Fig. \ref{sedpairs} also shows 
the \texttt{T400-Peak} spectrum (i.e., extracted around the maximum of the $\gamma$-ray flare;
see Table \ref{tab:deabs300} and  Fig. \ref{fig:fluxindex500}).
This spectrum unfortunately lacks X-ray coverage (the {\chandra} pointing started a few minutes later),
but it corresponds to the brightest $\gamma$-ray emission ever recorded from PKS\,2155--304.

Both synchrotron and IC peaks do not shift across 
the observed passbands, despite the large flux variations.
There is no evidence of the dramatic changes displayed by Mkn\,501 or 1ES\,1959+650.
Only at the flare maximum does the Compton peak become  visible in the
observed passband (between $\sim$400 and $\sim$600 GeV, depending on the EBL normalization).
The X-ray spectrum hardens apparently pivoting around the UV band. 
This behaviour is typically observed also in Mkn\,421, Mkn\,501 and 
1ES\,1959+650, but in these sources the amplitude of the spectral hardening
tends to be significantly more pronounced, leading to a shift 
of the synchrotron peak  in the hard X-ray band.
In the VHE band instead the behaviour is more complex,
since the  hardening at higher fluxes is also accompanied 
by a stronger curvature/cutoff  (see Fig. \ref{sedpairs}, right panel).

The slopes of the X-ray (above $\sim$1 keV)  and VHE  spectra are very similar, 
going from $\Gamma\sim$2.6 ($\pm \leq0.04$) to 2.9 ($\pm \leq0.11$)  between the high and low states.
It is also interesting to note that the X-ray spectrum is 
significantly less curved than the spectrum  measured by BeppoSAX
during the high state of 1998 \citep{chiappetti99}.
While the slopes above few keV are quite similar (see Fig. \ref{sedpairs}),
the flare in this night seems characterized by a higher luminosity in the soft X-ray band.

\subsection{Compton dominance}
The time evolution 
of the \nufnu fluxes close to the SED peaks
is provided by Fig. \ref{nufnu}.
As one can see, the $L_{\rm C}$/$L_{\rm S}$ ratio is of the order of $\sim$8, 
but it is rapidly variable -- on the same timescales as the flux variations --
decreasing  to the usual values of $\lesssim$1 in a few hours.

We estimated the Compton dominance using also the more detailed
spectral shape obtained for the \texttt{T300-High} dataset,  
and renormalizing the flux  to the 2 and 4-minute light curves. 
Considering the integrated luminosity over a decade in energy, 
namely 0.3--3 TeV and  0.3--3 keV, the $L_{\rm VHE}$/$L_{\rm X}$ ratio 
is  $\approx$10 in the first minutes of the simultaneous window. 
Assuming that the cubic trend between $\gamma$-ray and X-ray fluxes is 
maintained  up to the flare maximum 
(which occurred $\sim$25 minutes before the start of the X-ray observation),
one obtains $L_{\rm VHE}$/$L_{\rm X}$$\approx$14 at the flare maximum.
The \nufnu flux \emph{at the peak of the SED} can instead be estimated from
the log-parabolic fit, based on the assumption that the curvature of the spectrum
also remains the same  outside the observed energy band.  
Using the aforementioned procedure, 
one obtains a $L_{\rm C}$/$L_{\rm S}\sim$5 in the simultaneous window, 
increasing to $\sim$6 at the flare maximum.
The values are lower because 
the estimate of the synchrotron peak 
energy ($\sim$45 eV) locates the X-ray band farther away from the synchrotron peak 
than the VHE band is from the Compton peak ($\sim$400-600 GeV).

This is the first time that such high $L_{\rm C}$/$L_{\rm S}$ ratios are observed 
in an HBL, irrespective of the amount of the intergalactic EBL absorption.
While the Compton dominance can be up to 100 in powerful FSRQ,
boosted by the IC emission on the intense external photon fields from the disk 
and the BLR \citep{sikoraec94,gg98}, so far it has been of the order of unity or less 
in HBL.  
High Compton dominances in HBL were previously obtained only in the presence of
a very high density of the EBL \citep{hegra1426,nature_ebl}, which is now considered 
very unlikely \citep{nature_ebl,madausilk,franceschini}.

\subsection{Overall SED properties}
The overall SED  of PKS 2155--304 in the highest and lowest states during this night 
is shown in Fig. \ref{sedtot}, together with historical data.
To plot the $\gamma$-ray spectral shape in more detail
while preserving the amplitude variation, 
the VHE spectra shown in Fig. \ref{sedtot} are the \texttt{T300-High} and \texttt{T300-Low} average spectra
described in Sect 4.1 (see Fig. \ref{fig:fluxindex300} and \ref{fig:specs300}),
rescaled to match the highest and lowest fluxes in the 4-minute light curve.
Since within these two subsets, the spectra are compatible with a constant value 
(see Fig. \ref{fig:fluxindex300}), 
this procedure should not introduce significant distortions in the VHE spectral shape.
For clearer visibility, the SED focuses on data above $\sim$$10^{11}$ Hz,
thus excluding the radio frequencies (VLBI range). 
With the flux and variability timescale ($\lesssim$3 hrs) 
shown at optical (and higher) frequencies, 
the synchrotron emission coming from this region becomes self-absorbed 
already at frequencies below $\sim$$10^{12}$ Hz, and is thus not expected 
to contribute significantly in the radio range.

From the results of the log-parabolic fits, 
it is possible to estimate the peak frequency and luminosity 
of both the synchrotron and Compton emission. 
By considering the highest state, the synchrotron and Compton peak luminosities 
are estimated to be at 1.5$\times10^{46}$ and 7$\times10^{46}$ erg/s, respectively, 
with peak frequencies of $\sim$50 eV and $\sim$260 GeV.
Extrapolating the X-ray data with
the observed trends up to the flare maximum,
one obtains 1.6$\times10^{46}$ and 9$\times10^{46}$ erg/s at 
$\sim$50 eV and $\sim$500 GeV, respectively.
In the lowest state, instead, the peak luminosities 
become comparable
(1.2$\times10^{46}$ and 1.1$\times10^{46}$ erg/s, respectively, 
at  $\sim$50 eV and $\sim$70 GeV).
As discussed in Sect. 4.1, even the lowest possible EBL density 
does not alter substantially these estimates.

In the optical band, the contribution of the host galaxy is 
negligible: the host galaxy is resolved in optical 
\citep{falomo91,falomo96} and NIR \citep{kotilainen98}, 
and found to be a giant elliptical galaxy of M(R)=-24.4. 
This translates into an apparent m(V)$\simeq15.7-15.8$ using the typical 
colours for an elliptical galaxy \citep[V-R=0.61-0.71,][]{fukugita95}.
The optical flux of PKS 2155--304 is thus always dominated by the jet emission
\citep[see e.g.,][]{dolcini07}:
the historical light curves from long-term photometric monitoring in the V band 
show variations in the range 12.3--13.9 magnitudes \citep{carini92, osterman07}.
During this night, the optical flux is very high but still far from 
the highest fluxes observed from this object.
 
In the X-ray band, instead, the flux 
is close to the highest state observed historically, as measured by  
\rxte-PCA in 1996  \citep[$F_{2-10} \sim$1.6\ergs{-10},][]{rxte96}.
PKS\,2155--304 is  observed frequently in the X-ray band, since it is a calibration 
and monitoring source for {\xmm}, {\chandra}, and {\swift}, 
but was never found at the level observed during this night, 
in the 2--10 keV band \citep[see e.g.][]{distance,massaro2008}.
Quite interestingly, the {\chandra} spectrum in the highest state also seems to connect
smoothly with the flux and spectrum measured by \rxte-HEXTE in 1996 
\citep[][butterfly in Fig. \ref{sedtot}]{rxte96}.
It is also interesting to note that the extrapolation to lower energies of the 
log-parabolic fit for the average \texttt{T200} spectrum matches quite well the flux 
in the optical band.

\section{Summary of the main observational findings}
Before discussing the implications in the context of  
blazar physics, it is useful to  summarize 
the main observational findings of this 
phenomenologically rich dataset.

\begin{itemize}

\item Large-amplitude $\gamma$-ray variations are accompanied by small
X-ray and optical changes. In a few hours, the $\gamma$-ray flux changes 
by more than an order of magnitude, reaching a luminosity of $\approx10^{47}$ erg/s,
while the X-ray flux varies by only a factor $\sim$2 overall, 
and the optical V flux by less than 15\%.

\item The X-ray and $\gamma$-ray emission correlate strongly, 
overall and on short (sub-hour) timescales.
On very short timescales (few minutes), 
the behaviour might be more complex.

\item There is no evidence of time lags between the X-ray and $\gamma$-ray emissions,
with a 95\% upper limit of $\sim$3 minutes for the overall light curve.
In addition, no lags are found between the hard and soft energies of each passband.

\item The optical light curve shows a $\sim$15\% rise that appears to start
simultaneously with the $\gamma$-ray flare, but develops on much longer timescales,
reaching a plateau about 2 hours later than the VHE peak.
The optical emission does not show any correlation with the other two bands
on short timescales.

\item When correlated, \emph{the $\gamma$-ray flux decreases as the cube of
the X-ray flux} ($F_{\rm VHE}\propto F_{\rm X}^{\sim3}$).
This cubic relation holds both during the overall decaying phase
and considering shorter intervals separately.

\item The X-ray and $\gamma$-ray ($>300$ GeV) \emph{spectra correlate as well}, 
following similar patterns in their time evolution, but again of different amplitudes.
The VHE variations ($\Delta\Gamma_{\rm VHE}=0.65$) are wider by a factor of 3 than those 
in the X-ray band ($\Delta\Gamma_{\rm X}=0.21$).

\item  The $\gamma$-ray spectra are significantly curved, and
\emph{the curvature changes with time}, in correlation with the flux state:
the higher the flux, the more curved the spectrum
(the curvature parameter $b$ goes from $1.2\pm0.2$ 
at the maximum to $0.38\pm0.14$ in the low state).

\item The synchrotron and Compton peaks show no strong shift in frequency
across the observed bands, despite the dramatic luminosity 
changes ($100\times$ the quiescent state, 
$20\times$ in this single night), remaining close to their historical values. 
From the curvature of the spectra, the synchrotron peak can be estimated 
at $\approx40-50$ eV constantly along the night,
while the Compton peak shifts from $\approx500\pm100$ GeV at the maximum 
to $\approx70^{+110}_{-10}$ GeV at the end of the night 
(uncertainties given by the range on the EBL normalization).

\item \emph{A very large Compton dominance} is observed ($L_{\rm C}/L_{\rm S}$$\gtrsim$8). 
This is the first time that such a high $L_{\rm C}/L_{\rm S}$ 
ratio is seen in an HBL, irrespective of the level of intergalactic EBL absorption.
However, it also evolves rapidly, decreasing in a few hours 
to the more usual values of $\lesssim$1.

\item X-ray and VHE spectra shows a similar ``harder-when-brighter'' behaviour 
in the simultaneous window. The X-ray data alone also sample the start 
of another flare, characterized instead by a ``softer when brighter''  behaviour.

\end{itemize}

\section{Discussion}
The  results obtained from this campaign 
seem to  both corroborate and challenge the one-zone SSC interpretation
at the same time.

On the one hand, the strong correlation between variations in the X-ray and VHE bands 
-- the emission follow the same variability 
patterns in terms of both flux and spectrum, and without apparent lags --  
do indicate that the same particle distribution, in the same physical region,
is likely responsible for the activity in both energy bands.
The simultaneity of the occurrence of the  optical and VHE flares  
also suggests that the emission 
in all three bands is responding to the same flaring event. 

On the other hand, the cubic relation during the decaying phase
cannot be easily accounted for 
within a one-zone SSC scenario.
This  is shown for example by \citet{katar05}, 
who investigated the different X-ray/VHE correlations 
achievable in a one-zone SSC scenario, 
for a plausible range of physical parameters
and considering all possible combinations of fluxes from both sides of the SED peaks.
The relation between X-ray and $\gamma$-ray fluxes 
when both bands sample the emission above their respective SED peaks
is typically less than quadratic. Even imposing the Thomson condition
for the IC scattering of X-ray photons, which in this case would require
the extreme values of $\delta\sim$800 and $B\sim$0.6 mG (see next par.),
at most a quadratic relationship can be obtained, 
but not in the decaying phase if the latter is due to radiative cooling.
In general, correlations steeper than quadratic (cubic or even more)
are indeed possible, but only when the X-ray band corresponds to
frequencies below or close to the synchrotron peak, as in the case of the 1997 flare of Mkn\,501 
\citep[see e.g.,][]{tav501}.

A further ingredient is therefore needed to obtain
a cubic relation, in a one-zone SSC scenario. 
It is important to remind here that ``one-zone''
does not mean the request that the entire SED is produced by a single zone.
It is generally understood that the SED is always formed by the superposition
of radiation from different zones along the jet, both in space 
(e.g., the radio emission has to come from a much larger region than for
the rapidly variable X-ray or optical emission, to avoid synchrotron self-absorption) 
and in time (multiple injections).
What is generally tested with a ``one-zone scenario'' is the hypothesis that 
at any given time -- but during large flares in particular -- 
it is one single region or population
that dominates the entire radiative output, determining both peaks of the SED.
This scenario has been very successful in  explaining so far the 
spectral and temporal properties of blazars, both
among different objects and in single sources during large flares,
(e.g., Mkn\,501, 1ES\,1959+650).
We now consider if this can also be the case for this flare of PKS\,2155--304.

\begin{figure}[t]
\centering
   \includegraphics[width=9cm]{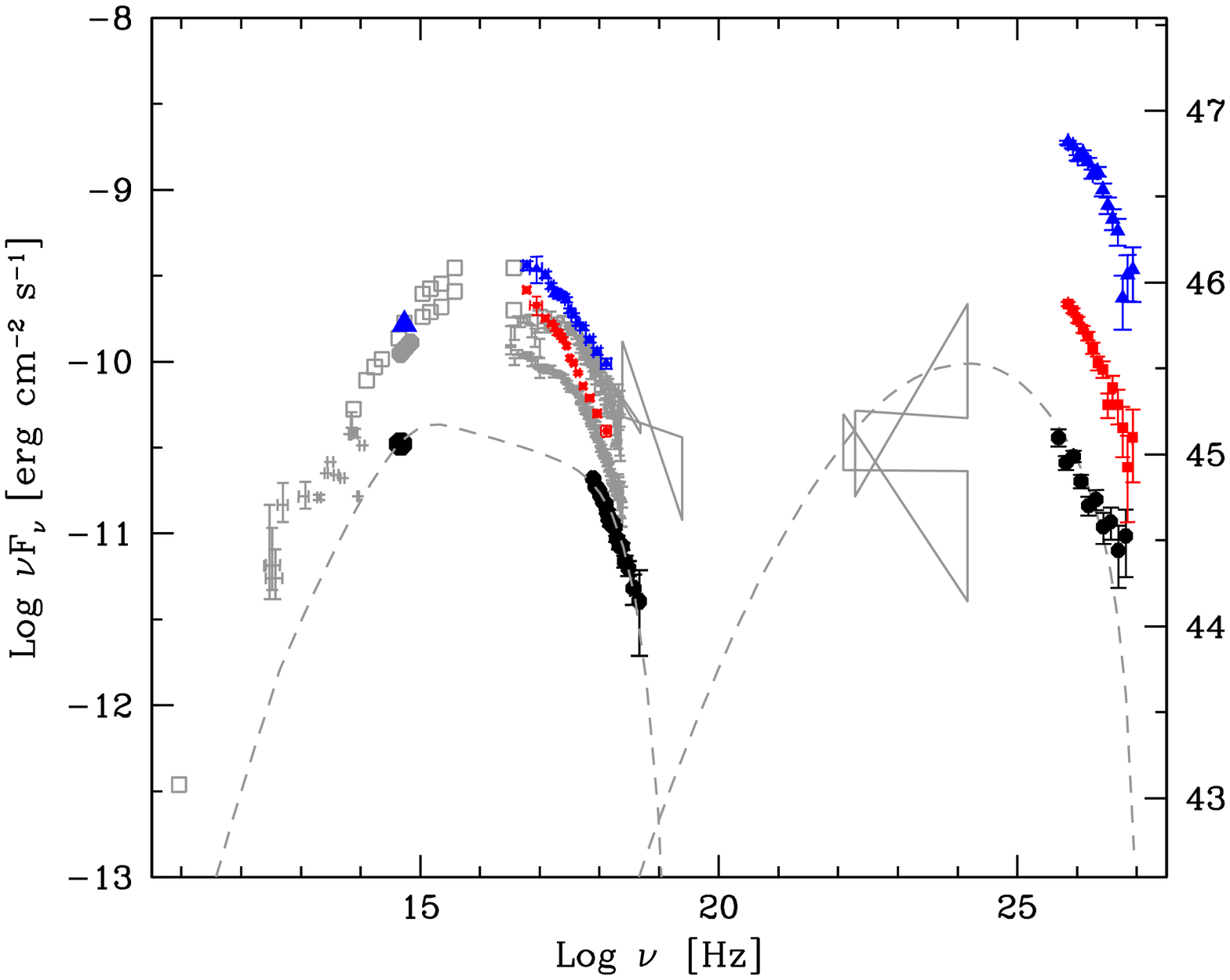}
     \caption{Synthetic SED  of PKS\,2155--304 showing the highest and lowest simultaneous 
     states during this night, together with historical data 
     \citep[shown in grey, see][and references  therein]{chiappetti99,iso,2155mwl}.  
     The hard X-ray data (butterfly) corresponds to the \rxte-HEXTE spectrum 
     in the high state of 1996 \citep{rxte96}. 
     The right axis gives the luminosity scale in erg/s. 
     Highest state (blue triangles): \texttt{T300-High} spectrum
     scaled to the highest VHE flux in the 4-min light curve. 
     Lowest state (red squares):  \texttt{T300-Low} spectrum
     scaled to the lowest VHE flux in the 4-min light curve. 
     The X-ray states in the corresponding time bins are practically equal  
     to the \texttt{T300-Xmax} and \texttt{T400-Xmin} spectra,
     which are thus plotted here. 
     The dashed line shows the one-zone SSC fit of the 2003 multiwavelength campaign 
     \citep[black circles,][]{2155mwl}. }
     \label{sedtot}
\end{figure}


\subsection{One-zone SSC analysis}
\subsubsection{Thomson condition and beaming factor}
In a single-zone, homogeneous SSC model, 
the knowledge of the frequencies of the synchrotron and Compton SED peaks 
($\nu_s$ and $\nu_c$, respectively), their luminosities ($L_s$ and $L_c$),
and the requirement of Thomson condition for X-ray photons ($\nu_x$) 
form a closed system of three equations in the unknown variables
$\delta$ (beaming factor), $B$ (magnetic field), and $R$ (size) of the emitting region
\citep[see e.g.,][ Appendix A]{katar05}.
These values can then be compared  with the 
size of the emitting region obtained from variability timescales 
($R\leq c t_{var} \delta (1+z)^{-1}$), and with the condition that
the cooling times of the electrons emitting at the peak ($\gamma_{peak}$) 
is equal to or shorter  than the variability timescales 
\citep[namely, $\gamma_{cool}\leq\gamma_{peak}$; see e.g.,][]{tavecchio98}.

With the parameters measured for this flare ($\nu_s\simeq50$ eV, $\nu_c\simeq0.5$ TeV,
$L_s=1.6\times10^{46}$ erg/s, $L_c=1\times10^{47}$ erg/s), 
the Thomson limit for photons in the middle of the {\chandra} passband  ($\nu_x=1$ keV)
requires that $\delta\sim$800, $B\sim$0.6 mG and $R\sim$1$\times10^{15}$ cm.
This solution seems unrealistic: besides the quite extreme values
of the beaming factor and severe efficiency problems, 
it is  not consistent with the VHE spectra.
The Thomson $\gamma$-ray spectrum  should extend to 10 TeV 
with the same slope shown by the X-ray spectrum up to 1 keV, in contrast  
to observations. 

For electrons emitting at the SED peaks, instead,
PKS\,2155--304 is almost certainly in the Thomson regime. 
The Thomson regime is already satisfied for $\delta\approx15$, but 
the condition of transparency for 1-TeV photons \citep[see e.g.,][]{dondi95,fabian}
requires $\delta\gtrsim30$, adopting for reference the variability timescale of the
main $\gamma$-ray flare ($t_{var}\sim$1 hr).
The Thomson regime for the SED peaks is also indicated by the absence of a significant
change of $\nu_s$ ($<1.5\times$) in the presence of an inferred 
$\sim$3$\times$ shift of $\nu_c$. 
Since in the KN regime 
the ratio $\nu_c/\nu_s\propto\gamma_{peak}^{-1}\propto \nu_s^{-1/2}$, 
the change in $\nu_c$ should have been accompanied 
by an increase in $\nu_s$ by 9$\times$ (vs. $\sqrt{3}\times$ in Thomson),
bringing it well within the {\chandra} band. 

Therefore, the SED peaks of PKS\,2155--304 are in the Thomson regime, 
but a part of the {\chandra} passband is most likely affected by KN effects.
With $\delta\gtrsim$100,  consistency between the size of the emitting region
estimated from both the observed Compton dominance and the variability timescale is obtained,
yielding $R\approx$$1\times10^{16}$ cm, $B\lesssim$5 mG and a Thomson condition
at $\nu_x\lesssim0.3$ keV.  
This is also the result obtained by an analysis on the $B-\delta$ plane
following the analytical approach described in  \citet{tavecchio98}.

However, this solution is not consistent with the cooling times of the peak electrons 
being equal to or shorter than the escape timescales.
Because of the low values of the magnetic field and high beaming factors,
in all the above cases the energy of the electrons cooling within the variability timescale
($t^\prime=t_{var}\delta/(1+z)$), 
either by synchrotron or IC, is extremely high, namely $\gamma_{cool}>6\times10^{6}$. 
This value is well above $\gamma_{peak}\sim$9$\times10^4$, meaning that the whole
electron distribution up to the energies sampled by our observations 
did not have time to cool.

In the one-zone frameset, therefore, the peaks of the SED cannot be explained by
radiative cooling.  This result is quite different from all previous estimates 
for the SSC parameters in PKS\,2155--304, and is mainly due to the increase 
by a factor $\sim$10 in the separation  between the synchrotron and Compton peak frequencies.

\subsubsection{Expanding blob}
An intriguing possibility  
is to explain  the observed decrease of both synchrotron and IC emission
as adiabatic cooling due to a rapid expansion 
of the emitting region from an initially very compact size (``explosion''),
with the total number of particles $N$ roughly constant. 
This hypothesis has two immediate advantages:
1) it accounts for the initial high value of the Compton dominance as well as its
rapid decrease 
\citep[since in the Thomson limit $L_c/L_s=U^\prime_{\rm rad}/U_{\rm B} \propto N R^{-2}\langle \gamma^2 \rangle$,
where $\langle \gamma^2 \rangle$ is the average over the electron distribution,
see e.g.][]{ggsequence2};
2) it can explain in a natural way the change of curvature of the VHE spectra
through a change of internal transparency to $\gamma$-$\gamma$ interactions.

The initial region cannot be too compact, or otherwise
it would not be transparent to $\gamma$-ray photons. 
However, in the case of PKS\,2155--304, the steep and curved VHE spectra
can allow for a moderate amount of possible internal absorption to be 
considered  ($\tau_{\gamma\gamma}\sim$1--2, as obtained for example with $\delta\sim$30--40),
without requiring anomalous spectral shapes.
The optical depth $\tau$  increases with $\gamma$-ray photon energy as 
$\tau(\nu_{\gamma})=\tau_0 (\nu_{\gamma}/\nu_0)^{\alpha}$,
where $\alpha$ is the energy spectral index of the synchrotron target photons 
($F(\nu)\propto \nu^{-\alpha}$; $\alpha\equiv \Gamma-1$), 
and has a radial dependence $\propto R^{-2}$, for a constant $N$.
Therefore, internal absorption can explain the stronger curvature
in the brightest state  and a rapid expansion would decrease the optical depth  
making the $\gamma$-ray spectrum less curved at lower fluxes, as observed. 
In fact, the observed change of curvature is opposite to that  expected 
from radiative cooling \citep[e.g.,][]{massaro2008}, though it could 
also be due to a change of the maximum electron energy.

To obtain the cubic correlation, however, one must assume correlated variations 
of a second parameter, in order to shift the radiative output 
from the IC to the synchrotron channel. 
This can be achieved by varying the magnetic field $B$ \citep[see e.g.][]{coppiahar99}:
for a given particle injection, an increase of the magnetic field can lead to a reduction 
($\propto B^{-2}$)
of the flux of the inverse Compton $\gamma$-rays while increasing the synchrotron emission.

Using the same formalism  as in \citet{katar05}, 
it is possible to find the relation that yields  $F_{\gamma}\propto F_x^3$.
The evolution of the radius $R$ and magnetic field $B$
can be parameterized as $R=R_0 (t/t_0)^r$ and $B=B_0 (t/t_0)^{-m}$ 
\citep[see also][]{atoyan99}. The adiabatic losses sustained by the particles during 
the expansion are taken into account.
With the values for the spectral indices before and after the synchrotron peak
of $\alpha_1=$0.7 (assumed for the optical spectrum) and $\alpha_2=$1.4
(as measured in the X-ray band), respectively,  
a cubic correlation is obtained with $m/r\simeq-0.4$ 
( $\pm0.1$ considering $\alpha_1$ between 0.5 to 0.9).  
Namely,  the magnetic field \emph{must increase as the blob expands 
at a rate} $B\propto R^{-m/r}=R^{+0.4}$, and on the same timescales as the flux variations.
This implies that the total energy of the magnetic field 
in the region must increase substantially, as $W_B\propto R^{+3.8}$. 
Thus it should be created  either locally (presumably by turbulent dynamo effects) 
or supplied from outside \citep[see e.g., discussion in][]{atoyan99}. 


However, such strong and rapid amplification of the magnetic field 
seems not consistent with the optical data.
Rapid variations of $B$ within the emitting region are also bound to
affect the synchrotron emission before the peak, produced by particles not yet cooled.  
With the previous parameters,  the observed decay of a factor 2 in the X-ray band 
would cause a decrease of the optical flux by $\sim$15\%
\citep[compare $F_s^2(t)$ vs. $F_s^1(t)$ in][]{katar05},
and on the same timescales as the X-ray and VHE variations.   
This disagrees with the optical light curve, 
which is still slightly rising (by $\sim$5\%) and then remains
basically constant during the overall X-ray/VHE decay
(see Fig. \ref{fluxgo}).
While it might be possible to further trim this scenario
to cancel out any variation in the optical band, 
this would require an extreme fine-tuning.

For all of these reasons, it is very unlikely that both of the peaks in the 
PKS\,2155--304 SED  during this night are produced 
by the synchrotron and SSC emission of one single zone.

\begin{figure}
\centering
   \includegraphics[width=9cm]{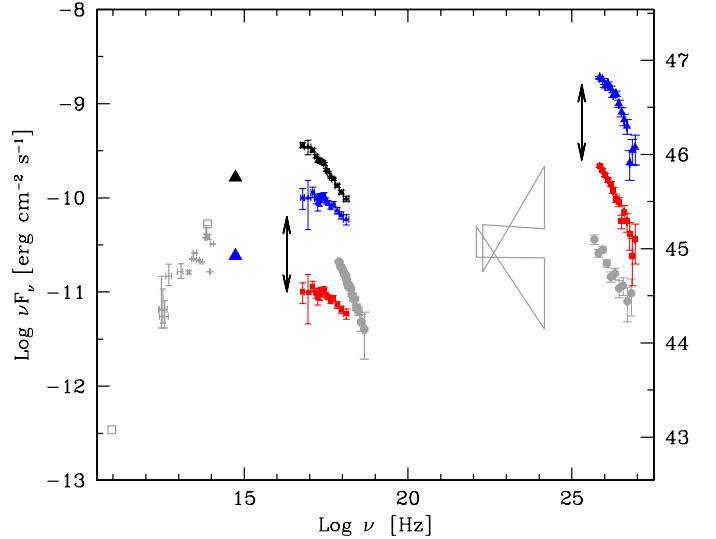}
     \caption{Possible SED of the new flaring component in PKS\,2155--304,
     which would vary ``up and down" beneath the  persistent emission of PKS\,2155--304 
     (see text).
     Blue triangles: SED of the flaring component at the flux maximum 
     inside the X-ray/VHE simultaneous window.
     Red squares: expected level of the X-ray flux assuming a 
     linear relation between X-ray and VHE flux variations.
     For reference, the same {\chandra} \texttt{T300-Xmax} spectrum and 
     optical data shown in Fig. \ref{sedtot} are plotted as well (highest black points).
     }
     \label{newcomp}
\end{figure}

\subsection{Two-SED scenario}
A simpler and possibly more realistic explanation is provided
by the superposition of two SEDs, produced by two different emitting zones.
The first is responsible for the usual ``persistent''  SED of PKS 2155--304,
which, presumably, peaks in the UV and at a few tens GeV, but with rather low VHE fluxes.
It is typically variable on longer timescales ($\sim$0.2--1 days),
as shown for example by the long {\it ASCA}, {\sax}, and {\xmm}  observations
\citep{tanihata,zhang02,zhang06}.
The second zone is responsible for the dramatic flaring activity 
of this night, and most likely of all the active period of July-August 2006.
It needs to be much more compact and have higher bulk motion
to account for the much faster variability timescales (0.05--1 hrs).
%
The true variations of the X-ray emission from this second zone can 
therefore be as large as the $\gamma$-ray ones,
but they are simply seen \emph{diluted} in the ``persistent'' component,
which have comparable or higher synchrotron fluxes.
These variations are instead  fully visible in the VHE band 
because there the contribution of the standard SED is at
much lower fluxes.

A two-zone scenario is rather common for explaining  major flares in blazars.
The main novelty of this event with respect to all previous HBL flares 
is that the bulk of the luminosity of the new component 
is now emitted in the Compton channel instead of the synchrotron channel.
The synchrotron emission of this flare 
is not bright enough to ``break through'' the  persistent emission 
and thus dominate both peaks of the SED.

Within this scenario, one can try to estimate the SED 
of this new component at its brightest state during the simultaneous observations,
assuming that the flux of the persistent component is roughly constant.  
This is shown in Fig. \ref{newcomp}.
In the $\gamma$-ray domain, the flux of the persistent component 
is supposed to be negligible with respect to the observed flare. 
The spectrum of the flaring region is therefore given, to a first order,
by the highest VHE state directly observed (the same as in Fig. \ref{sedtot}).
In the optical band,  at the maximum of the flare the detected variation 
is $\sim$15\%, so the flux in the flaring zone should be $0.15$ 
of the observed flux. 
In the X-ray band, we have used the lowest-state spectrum
observed at the end of the night as background file for the highest-state spectrum.
As a result, a spectrum with flux $F_{0.5-5}=1.98$\ergs{-10} is obtained,
which is well fitted by a log-parabolic model with $\Gamma_{1 \rm keV}=2.17\pm0.06$ 
and curvature $b=0.22\pm0.11$. 
This gives the estimate for the synchrotron 
peak position around 0.4 keV, which is also consistent with the optical flux being 
much lower than the X-ray flux (see Fig. \ref{newcomp}).

This underlying SED  yields comfortable values for the synchrotron-Compton modeling, 
and becomes consistent with the peak being due to radiative cooling.
Such scenario 
also agrees with the spectral variations being larger 
in the VHE band than in the X-ray band (for the same diluting effect), 
and would naturally account for the behaviour of the optical data as well,
if the optical rise is indeed associated with the VHE flare.
The optical band would be reacting to the electron injection in the rising phase
but not in the decaying phase, due to the much longer cooling times.
The variations at VHE 
can then be allowed to scale linearly or quadratically 
with the X-ray variations, as required by the specific modeling.

\subsection{Diagnostic of the flaring region}
Although the amplitude of the synchrotron variations of
the new component is unknown, 
important constraints can still be derived on its properties
in  both cases of a quadratic or linear 
relationship between  X-ray and $\gamma$-ray fluxes.

With the new parameters, a quadratic decay obtained through radiative cooling
is readily possible with a pure SSC scenario 
because the peak of the synchrotron emission falls now directly within the {\chandra} passband.
Indeed a solution in the Thomson regime can be obtained with the 
comfortable values of  $\delta\gtrsim30$, $B\sim$1 Gauss, 
and $\gamma_{peak}\simeq 3\times10^{4}$.  
However, the high Compton dominance requires
the emitting region to be very compact, with $R\simeq3-5\times10^{14}$ cm.
The flare therefore would originate in a region whose size is  
of the same order as the Schwarzschild radius  of the putative central black hole,
which is estimated to be $\sim$$10^{9} M_\odot$ \citep[see discussion in][]{bigflare}.

If instead the X-ray flux scaled linearly with the VHE flux,
it implies that the new component has a high Compton dominance ($\approx20$) 
constant in time, during all flux variations.
This behaviour points towards 
an origin of the $\gamma$-ray peak from  
external Compton emission rather than a pure SSC mechanism.

This is indeed expected in scenarios with a strong radiative interplay
between different parts of the jet, such as the ``Needle/Jet'' model proposed  
by \citet{ggrapid}: a compact and fast region moving throughout 
a larger jet sees the dense target field produced by the latter 
(which is thought to be responsible for the persistent SED), boosting significantly its Compton emission.
Unlike the way it is envisaged by the authors, however, 
the SED of such needle should also contribute significantly in the X-ray band,
in order to explain the small-amplitude but correlated variability.
Alternatively, the dominant seed photons could come from  
regions further down inside the jet, where the flow  has already strongly decelerated
\citep[i.e., at the VLBI scale,][]{markos03,piner}, or by circumnuclear radiation fields
if the flaring region is close to the central engine.
In the latter case, $\gamma$-$\gamma$ interactions with these photon fields 
could leave a distinct absorption feature imprinted onto the broad-band 
$\gamma$-ray spectrum \citep{akc08},
which the combined observations with \glast~ and IACT might be able to 
reveal \citep{luigi_glast}.
More complex models with several interacting zones or jet stratification 
have also been developed, and recently applied to the variability 
of PKS\,2155--304 in 2006 \citep[see e.g.][]{gilles08,meudon08}.
They seems capable of reproducing some features of the present observations
\citep[e.g.,][]{gilles08}.

\subsection{New mode of flaring in HBL}
On a more general basis,
it is interesting to compare this flare of PKS\,2155--304 
with the other major flaring events observed in both X-ray and VHE bands,
namely from Mkn\,421 in 2000 \citep{fossati08},  
Mkn\,501 in 1997 \citep{tav501,501} and 1ES\,1959+650 in 2002 \citep{1959}. 

There are many common traits:
they all display a pre-flare, ``persistent'' SED with the synchrotron emission
peaking in the UV/soft-X-ray band, a steep X-ray spectrum that hardens
during the flare pivoting around the UV/soft-X-ray band, and a 
flare luminosity approximately one order of magnitude higher than the 
typical source luminosity.  The mechanisms and properties 
of the flare injection thus seem common. 
It is the radiative output that now differs significantly.

In the previous events, such a high luminosity
was emitted mostly through  the synchrotron process, 
leading to a dramatic shift of the peak position in the SED 
according to the new peak frequency of the emerging component. 
The typical Compton luminosity,  even at the flare maximum, 
has always been equal to or less than the synchrotron power
(using the same EBL model for all sources).
In this event, instead, the bulk of the flare luminosity is emitted 
through the Compton channel, yielding only minor modifications 
of the overall synchrotron emission.

A bimodality therefore seems to emerge in the mode of flaring for HBL:
either synchrotron dominated or Compton dominated, with the most extreme example 
possibly being provided by the ``orphan flare'' event of 1ES\,1959+650,  
which likewise occurred during one single night (June 4, 2002).

It is intriguing to note that this difference might simply depend on the environment, namely
on the location of the flaring zone with respect to the region responsible 
for the persistent SED:  if the new injection/flare is taking place far away, 
there is little radiative interplay between the two zones, leading to a typical SSC-type 
flare. When the flare occurs close to it, or close to the black hole where external fields are 
more intense, the outcome is external Compton-dominated flares.
More campaigns targeted on the intra-night variability are needed to address this issue,
but the diagnostic potential on the jet structure and the location of the 
``$\gamma$-ray zone'' is very promising.


\section{Summary and conclusions}
A full night of simultaneous, uninterrupted observations 
in the optical, X-ray and VHE bands was performed during an exceptionally bright state of 
PKS\,2155--304 in July 2006.  A sampling of both light curves and spectra with 
unprecedented detail is obtained.  
For the first time among HBL, a high Compton dominance is observed,
with a peak luminosity reaching $\sim$$10^{47}$ erg/s.
The variations in the X-ray and VHE bands are confirmed to be highly correlated, 
both in flux and spectrally, but they follow a cubic relationship during the decay phase.
Homogeneous one-zone SSC scenarios do not provide a consistent explanation
for both the observed SED and the variability behaviour,
indicating that a single particle population cannot be responsible 
for both peaks of the SED during this night.

We have interpreted the data as the emergence of a new component in the SED,
strongly Compton-dominated and thus without enough luminosity
in the synchrotron channel to overcome the ``persistent'' emission of PKS\,2155--304. 
This new component must be either very compact
-- of the order of the Schwarzschild radius of the putative 
central black hole ($\sim$$10^9$ M$_{\odot}$)-- if the emission mechanism is pure SSC,
or external-Compton dominated, as would be expected in models with 
a strong radiative interplay between different parts of the jet. 

The richness and quality of the data obtained from this exceptional campaign 
provide a fundamental testbed for a time-dependent treatment of the emission scenarios,
which is needed to fully extract the information 
on the physical conditions in the source \citep[see e.g.,][]{coppiahar99,501,gilles08}.
This is beyond  the scope of this paper, and therefore left to future studies,
but it is a necessary step forward with respect to more common parametric fits, 
which simply postulate the prompt electron spectrum and treat different states
as uncorrelated emissions.

From an observational point of view,
the results from this night raise the bar for the requirements of future multiwavelength 
campaigns on HBL, at least  during major flaring events. 
With flares as rapid as a few minutes \citep[][]{bigflare}
and flux changes of 1 order of magnitude in less than an hour,
sufficiently long, continuous and strictly simultaneous observations are mandatory
for a meaningful analysis. 
Delays, gaps, or mismatches in the simultaneous coverage, even as short as 30 minutes, 
can be fatal for the diagnostic potential. 
The multiwavelength coverage should also care to sample the SED
on both sides of the emission peaks: 
as shown in the present case, without the optical light curve some scenarios 
could not be excluded.
In this respect, {\glast} will provide the missing piece of information 
on the $\gamma$-ray hump 
for many objects, though typically on longer timescales.
The all-sky monitor  provided by this observatory will be of great importance
to catch the  brightest Compton-dominated flares among HBL, 
triggering X-ray and IACT observations and thus allowing the study of this apparent flare bimodality 
(synchrotron or Compton-dominated) on a larger sample.

The flaring activity of PKS\,2155--304 in July 2006 
demonstrates in a compelling way that the fast, intra-night variability is 
of fundamental importance to gain insights into the acceleration mechanism, 
emission processes, and location and environment of the emitting regions,
which are simply lost or smeared out with a sparser, long-term sampling. 
The final goal is to achieve full multiwavelength coverage of 
large-amplitude and ultra-rapid events such as those observed on the night of July 27--28.

\begin{acknowledgements}
The support of the Namibian authorities and of the University of Namibia
in facilitating the construction and operation of \hess is gratefully
acknowledged, as is the support by the German Ministry for Education and
Research (BMBF), the Max Planck Society, the French Ministry for Research,
the CNRS-IN2P3 and the Astroparticle Interdisciplinary Programme of the
CNRS, the U.K. Science and Technology Facilities Council (STFC),
the IPNP of the Charles University, the Polish Ministry of Science and
Higher Education, the South African Department of
Science and Technology and National Research Foundation, and by the
University of Namibia. We appreciate the excellent work of the technical
support staff in Berlin, Durham, Hamburg, Heidelberg, Palaiseau, Paris,
Saclay, and in Namibia in the construction and operation of the
equipment. The authors thank the {\chandra} team for the help and support
during the prompt response to our ToO request, as well as the \rxte and 
\swift~ teams.
The authors wish to thank  P. Coppi, G. Ghisellini and F. Tavecchio for
the support during the initial phases of the project.
This research has made use of the NASA/IPAC Extragalactic Database (NED) 
which is operated by the Jet Propulsion Laboratory, 
California Institute of Technology, under contract with the National Aeronautics 
and Space Administration.
\end{acknowledgements}

\bibliographystyle{aa} 
\bibliography{chandranight} 
\end{document}